\documentclass{nature}

\usepackage{epsfig,amsmath}
\usepackage{graphicx}
\usepackage{color}
\usepackage[10pt]{moresize}
\usepackage{bm}
\usepackage{SIunits}

\let\oldcaption\caption
\renewcommand{\caption}{\sffamily \oldcaption}
\begin{document}

\title{Realization of ideal Weyl semimetal band in ultracold quantum gas with 3D Spin-Orbit coupling}

\author{Zong-Yao Wang$^{1,2,3,\dagger}$, Xiang-Can Cheng$^{1,2,3,\dagger}$, Bao-Zong Wang$^{1,4,5,\dagger}$, Jin-Yi Zhang$^{1,2,3,\dagger}$, Yue-Hui Lu$^{4,5}$, Chang-Rui Yi$^{1,2,3}$, Sen Niu$^{4,5}$, Youjin Deng$^{1,2,3}$, Xiong-Jun Liu$^{4,5\star}$, Shuai Chen$^{1,2,3,\star}$ and Jian-Wei Pan$^{1,2,3,\star}$}

\maketitle

\begin{affiliations}
\item Hefei National Laboratory for Physical Sciences at Microscale and Department of Modern Physics, University of Science and Technology of China, Hefei, Anhui 230026, China
\item Shanghai Branch, CAS Center for Excellence and Synergetic Innovation Center in Quantum Information and Quantum Physics, University of Science and Technology of China, Shanghai 201315, China
\item Shanghai Research Center for Quantum Science, Shanghai 201315, China
\item International Center for Quantum Materials, School of Physics, Peking University, Beijing 100871, China
\item Collaborative Innovation Center of Quantum Matter, Beijing 100871, China\\

$^\dagger$ These authors contribute equally to this work.

$^\ast$ To whom correspondence should be addressed. E-mail: xiongjunliu@pku.edu.cn (X.J.L.);
shuai@ustc.edu.cn (S.C.); pan@ustc.edu.cn (J.W.P.).
\end{affiliations}

\clearpage

\begin{abstract}
The Weyl semimetals \cite{hasan2017discovery,RMP2018,Wan2011,Burkov2011,lv2015experimental,TaAs} are three-dimensional (3D) gapless topological phases with Weyl cones in the bulk band, and host massless quasiparticles known as Weyl fermions which were theorized by Hermann Weyl in the last twenties~\cite{Weyl1929}. The lattice theory constrains that Weyl cones must come in pairs, with the minimal number of cones being two. The semimetal with only two Weyl cones is an ideal Weyl semimetal (IWSM) which is the optimal platform to explore broad Weyl physics but hard to engineer in solids.
Here, we report the experimental realization of the IWSM band by synthesising for the first time a 3D spin-orbit (SO) coupling for ultracold atoms. Engineering a 3D configuration-tunable optical Raman lattice~\cite{Theory_Liu}, we realize the Weyl type SO coupling for ultracold quantum gas, with which the IWSM band is achieved with controllability. The topological Weyl points are clearly measured via the virtual slicing imaging technique \cite{Theory_Liu,Gyu-Boong_Jo} in equilibrium, and further resolved in the quench dynamics, revealing the key information of the realized IWSM bands. The realization of the IWSM band opens an avenue to investigate various exotic phenomena based on the optimal Weyl semimetal platforms.
\end{abstract}

In a Weyl semimetal the valence and conduction bands meet at nodal points, where the quasiparticles are characterized by Weyl Hamiltonian and have linear dispersions~\cite{hasan2017discovery,RMP2018,Wan2011,Burkov2011,lv2015experimental,TaAs}. A Weyl node corresponds to a topological monopole, whose charge equals the Chern number of metallic Fermi surface enclosing the nodal point and defines the chirality of the Weyl fermions.
According to Nielsen-Ninomiya no-go theory \cite{No_go}, Weyl nodes emerge in pairs, with two nodes of each pair having opposite chirality, hence the minimal number of Weyl nodes in a semimetal is two. 
The semimetal with only two Weyl nodes is an ideal Weyl semimetal (IWSM)~\cite{IWSM1}, 
and is the most fundamental phase in the Weyl semimetal family.
As the two nodes in an IWSM cannot be trivially gapped out, any interacting phase born of IWSM is nontrivial.
Thus the IWSM can serve as a fertile ground to study not only noninteracting Weyl physics like chiral anomaly~\cite{Nielsen1983,Vishwanath2014}, but also exotic many-body phenomena, such as the space-time supersymmetry~\cite{Yao2015PRL} and non-Abelian chiral Majorana modes~\cite{Chan2017}, which may not be favored in the interacting Weyl semimetals with more Weyl points. So far various Weyl and Weyl-like phases have been widely reported, including the type-II Weyl semimetal~\cite{Soluyanov2015,LHuang2016}, triply degenerate semimetals~\cite{Bernevig2016,Ding2017}, and the magnetic Weyl semimetals~\cite{Magnetic1,Magnetic2,Morali2019}, while the IWSM is hard to engineer~\cite{IWSM3,IWSM4} and the direct observation is illusive.

Meanwhile, realization of novel topological models has been an active pursuit in ultracold atoms~\cite{Bloch2013PRL,Miyake2013PRL,2D_Esslinger,2D_Bloch,Wu_Science,Bloch2018,Song:2017uf}. Especially, the ultracold quantum gases with synthetic SO interactions provide pristine platforms to investigate exotic topological phenomena. The SO interactions synthesized in different dimensions have distinct fundamental features. The 1D SO coupling corresponds to Abelian gauge potential~\cite{Liu2009,Lin2011,Zhang2012}, while the 2D SO couplings corresponds to non-Abelian gauge potentials, with the famous paradigms including 2D Dirac~\cite{Baozong} and Rashba~\cite{JZhang2016} types, of which the former has been actively studied for realizing 2D quantum anomalous Hall (QAH) models in optical Raman lattices~\cite{Wu_Science,Sun_2018A}. The 3D SO interaction, characterized by 3D non-Abelian gauge potential, is the essential ingredient to realize high-dimensional topological matter. In particular, the emergent Weyl Hamiltonian in the Weyl semimetal~\cite{Dubcek2015,Xu2015,WangPRA2016} describes a 3D Weyl type SO coupling, whose realization has been a long-standing challenge in the field of ultracold atoms.

Here we realize and detect the 3D SO coupling and IWSM band for ultracold $^{87}$Rb atoms based on the recent proposal~\cite{Theory_Liu}, with the Hamiltonian in the 3D Bloch momentum $\boldsymbol{q}$-space
\begin{equation}\label{WeylH}
  H_{\text{Weyl}}=\boldsymbol{h}(\boldsymbol{q})\cdot\boldsymbol{\sigma},
\end{equation}
where $\boldsymbol{\sigma}=(\sigma_{x},\sigma_{y},\sigma_{z})$ are the Pauli matrices. The Hamiltonian $H_{\text{Weyl}}$ at a fixed $q_z$ renders a 2D QAH model, whose topology is modulated by $q_z$. The number of Weyl points can be tuned by controlling two-photon detuning of Raman couplings. The Weyl points are clearly resolved by virtual slicing reconstruction imaging technique~\cite{Theory_Liu,Gyu-Boong_Jo} and also quench dynamics.

\section{Construction of 3D Spin-Orbit coupling}

The 3D SO coupling is constructed with 3D optical Raman lattices, as outlined in Fig.\ref{fig1}(a).
The spin is defined from the magnetic sublevels of $^{87}$Rb atoms $\mid\uparrow\rangle=\left|1,-1\right\rangle$ and $\mid\downarrow\rangle=\left|1,0\right\rangle$.
A bias magnetic field $\boldsymbol{B}=14.5$G provides the quantization in $y$-axis and the Zeeman splitting of $10.2\text{MHz}$.
The $|1,+1\rangle$ state is effectively excluded.
Three retro-reflective laser beams $\boldsymbol{E}_{i}(i=x,y,z)$ with wavelength $\lambda=787$nm ($k_{0}=2\pi/\lambda$ and recoil energy $E_{\rm{r}}=\hbar^{2}k_{0}^{2}/2m$) shine on the atoms in $i$-th direction, each having two orthogonal polarization components $E_{ij}(j=x,y,z)$.
$E_{xy}$, $E_{yx}$ have frequency $\omega_{1}$, $E_{xz}$, $E_{yz}$ have frequency $\omega_{2}$, and $E_{zy}$, $E_{zx}$ have frequency $\omega_{3}$.
We set $\omega_3-\omega_1\approx 2\pi\times 10.2\text{MHz}$, matching the Zeeman splitting of $\mid\uparrow\rangle$ and $\mid\downarrow\rangle$, while $\omega_1-\omega_2=2\pi\times 200\text{kHz}$.
Two $\lambda/4$ waveplates with optical axis aligned along $\hat{z}$ are placed in front of the retro-reflectors,
producing $\pi$ phase shift between $E_{xz(yz)}$ and $E_{xy(yx)}$.

Our realization is based on configuration-tunable lattices. $E_{xy}$ and $E_{yx}$ form a square lattice, together with $E_{xz}$ and $E_{yz}$, deforming into the checkerboard lattice in x-y plane $\mathcal{V}_{2D}(x,y)=V_{2D}\cos{k_{0}x}\cos{k_{0}y}$ upon adjusting relative phase between $\boldsymbol{E}_{x}$ and $\boldsymbol{E}_{y}$ to $\phi=0$ (See Methods).
The 2D lattice potential $\mathcal{V}_{2D}(x,y)$ can be rewritten as $\mathcal{V}_{2D}(u,v)=V_{2D}\left(\cos^{2}{k_1u}+\cos^{2}{k_1v}\right)$ along $\hat u$ and $\hat v$ directions, with $\hat{u}=(\hat{x}+\hat{y})/\sqrt{2}$, $\hat{v}=(\hat{x}-\hat{y})/\sqrt{2}$ and $k_1=k_{0}/\sqrt{2}$.
The density profile of $\mathcal{V}_{2D}$ is shown in Fig.\ref{fig1}(b).
The total 3D lattice potential $\mathcal{V}_{Latt}\left(u,v,z\right)=\mathcal{V}_{2D}(u,v)+\mathcal{V}_{z}(z)$, where
$\mathcal{V}_{z}(z)=V_{z}\cos^{2}{k_{0}z}$ along $z$ direction, with depth $V_{z}\propto\left|E_{zx}\right|^{2}+\left|E_{zy}\right|^{2}$.

Raman couplings are generated by beam pairs $\left(E_{zx},E_{xy}\right)$ and $\left(E_{zy},E_{yx}\right)$,
with $\Omega_{1}=\Omega_{zx}\cos{k_{0}z}\sin{k_{0}x}$ and $\Omega_{2}=\Omega_{zy}\cos{k_{0}z}\sin{k_{0}y}$, where Rabi frequencies $\Omega_{zx}\propto E^*_{zx}E_{xy}$ and $\Omega_{zy}\propto E^*_{zy}E_{yx}$ are set as $|\Omega_{zx}|=|\Omega_{zy}|=\Omega_{0}$.
The large frequency difference $\omega_2-\omega_1$ prevents the couplings between $(E_{zy}, E_{xz})$ and $(E_{zy}, E_{yz})$.
For the circular polarized $\boldsymbol{E}_{z}$, $\Omega_{1}$ and $\Omega_{2}$ have $\pi/2$ phase shift~\cite{Theory_Liu}.
Then the Raman couplings, having 2D forms in $(x,y,z)$ coordinates, take 3D forms in the deformed checkerboard lattice bases and $\Omega_{R}\left(u,v,z\right)=\Omega_{u}\left(u,v,z\right)\sigma_{x}+\Omega_{v}\left(u,v,z\right)\sigma_{y}$, where $\Omega_{u}\left(u,v,z\right)=\sqrt{2}\Omega_{0}\cos{k_{0}z}\sin{k_1u}\cos{k_1v}$ and $\Omega_{v}=\Omega_{u}\left(u\leftrightarrow v,z\right)$.
As shown in Fig.\ref{fig1}(b), $\Omega_{u}$ ($\Omega_{v}$) is antisymmetric along $\hat{u}$ ($\hat{v}$) and symmetric along $\hat{z}$, leading to spin-flipped hopping in $\hat{u}$ ($\hat{v}$) direction.
We finally reach the Hamiltonian as
\begin{equation}\label{3DSOC}
    H=\frac{\hbar^2\boldsymbol{k}^{2}}{2m}+\mathcal{V}_{Latt}(u,v,z)+\Omega_{u}(u,v,z)\sigma_{x}+\Omega_{v}(u,v,z)\sigma_{y}+\frac{\delta}{2}\sigma_{z},
\end{equation}
where $\hbar\boldsymbol{k}$ is momentum, $m$ is the mass of an atom and $\delta$ is the two-photon Raman detuning.

With the tight-binding (TB) approximation, Eq.(\ref{3DSOC}) in Bloch momentum space reads \cite{Theory_Liu,WangPRA2016}
\begin{equation}\label{TBH}
  \begin{split}
    H_{\text{TB}}(\boldsymbol{q})&=\boldsymbol{h}(\boldsymbol{q})\cdot\boldsymbol{\sigma}\\
    &=2t_\text{so}(\sin{q_{u}}\sigma_{x}+\sin{q_{v}}\sigma_{y})+\left[m_{z}-2t_{z}\cos{q_{z}}-2t_{1}\left(\cos{q_{u}}+\cos{q_{v}}\right)\right]\sigma_{z},
  \end{split}
\end{equation}
where $\boldsymbol{q}=(q_u, q_v, q_z)$ is dimensionless quasi-momenta, $t_{z,1}$ ($t_\text{so}$) denote the spin-conserved (spin-flipped) hopping coefficients, and $m_{z}=\delta/2$.
For each fixed $q_z$, $H_{\text{TB}}$ gives to a 2D QAH model in $u$-$v$ plane~\cite{Theory_Liu,Wu_Science,Baozong,Sun_2018A}, with topology modulated by both $m_Z$ and $q_z$. The Weyl points emerge when the Chern number of QAH models changes versus $q_z$.

The experiment starts from an optical trapped $^{87}\textrm{Rb}$ Bose-Einstein condensate (BEC) with $2.0\times10^{5}$ atoms in $\mid\uparrow\rangle=\left|1,-1\right\rangle$ state.
As the 3D Raman lattice beams are adiabatically ramped up in 100ms, the BEC is loaded into the ground state of Eq.(\ref{3DSOC}).
The probe beams are shining along $\hat{z}$ and $\hat{y}$ (shown in Fig.\ref{fig1}(a)) after the atoms are freely released for 22ms,
to take the spin-resolved time-of-flight (ToF) 2D images on CCD-Z and CCD-Y, as shown in Fig.\ref{fig1}(c).
The $\mid\uparrow\rangle$ clouds are from the diffraction of the 3D optical lattices.
On CCD-Z, the major portion of the BEC is at momentum $\left(k_{x},k_{y}\right)=(0,0)$ and four small dots at momenta $\left(k_{x},k_{y}\right)=(\pm k_0,\pm k_0)$.
On CCD-Y, there are two diffractions are at $(k_{x}, k_{z})=(0, \pm 2k_0)$ and the overlaps of the four diffractions at $(k_x=k_0, k_y=\pm k_0, k_z=0)$, $(k_x=-k_0, k_y=\pm k_0, k_z=0)$.
The $\mid\downarrow\rangle$ clouds are from the Raman couplings. 
On CCD-Z, they are at $\left(k_{x},k_{y}\right)=(\pm k_0,0)$ and $\left(k_{x},k_{y}\right)=(0,\pm k_0)$. On CCD-Y, they are at $\left(k_{x},k_{z}\right)=(k_0,\pm k_0)$, $\left(k_{x},k_{z}\right)=(0,\pm k_0)$ and $\left(k_{x},k_{z}\right)=(-k_0,\pm k_0)$.
The 3D momentum distributions of the $\mid\downarrow\rangle$ clouds, as illustrated in Fig.\ref{fig1}(d), confirm the successful realization of 3D SO coupling.

\section{Topological phase diagram and Weyl node measurement}

The Weyl node momentum $\boldsymbol{q}^{\text{W}}$ is determined via $\left|\boldsymbol{h}(\boldsymbol{q}^{\text{W}})\right|=0$.
Around these points, the linear expansion $H(\boldsymbol{q}^\text{W}+\delta\boldsymbol{q})=2t_\text{so}\delta q_{u}\sigma_{x}+2t_\text{so}\delta q_{v}\sigma_{y}+2t_{z}\sin{q_z^{\text{W}}}\delta q_{z}\sigma_{z}$ renders the Weyl Hamiltonian around the $\boldsymbol{q}^{\text{W}}$ node. 
Fig.\ref{fig2}(a) is the calculated topological phase diagram of Eq.(\ref{3DSOC}).
The regions with number of Weyl nodes from 0 to 8 are presented by colors.
Importantly, there are phases with only two Weyl nodes, corresponding to the IWSM phase.
The insets of Fig.\ref{fig2}(a) show the Weyl nodes in the 3D Brillouin zone, which marked by $\oplus$ or $\ominus$ chirality.

To determine the positions of Weyl nodes in experiment, one can characterize the Chern number of 2D band structures for given $q_{z}$ plane. 
In fact, Hamiltonian Eq.(\ref{3DSOC}) delineates a set of 2D QAH bands in $u$-$v$ planes, stacked along $q_{z}$.
The locations of Weyl nodes are characterized as the points where the topology of 2D QAH bands changes via scanning $q_{z}$~\cite{Theory_Liu}.

In the experiment, about $2.0\times10^{5}$ atoms with temperature around 150nK are adiabatically loaded into the lowest bands of Eq.(\ref{3DSOC}).
After a spin-resolved ToF imaging along $\hat{z}$ direction, we obtain the 2D momentum distribution ($q_u$ and $q_v$) of the atom in $\mid\uparrow\rangle$ ($n_{\uparrow}(\boldsymbol{q})$) and $\mid\downarrow\rangle$ ($n_{\downarrow}(\boldsymbol{q})$) on CCD-Z, with $q_{z}$ being integrated out.
Spin polarization at $\boldsymbol{q}$ is calculated by $P(\boldsymbol{q})=\left[n_{\uparrow}(\boldsymbol{q})-n_{\downarrow}(\boldsymbol{q})\right]/\left[n_{\uparrow}(\boldsymbol{q})+n_{\downarrow}(\boldsymbol{q})\right]$.
As shown in Ref.~\cite{Theory_Liu}, the Weyl band has an emergent magnetic group symmetry, for which we can reconstruct the 3D topological bands from a series of $q_z$-integrated 2D spin textures.
In particular, for a given Raman detuning $\delta_0$ in Eq.(\ref{3DSOC}), the observed spin texture in $q_u$-$q_v$ plane on CCD-Z, with $q_z$ being integrated out, is identical to its 2D spin texture on the $q_z=\pi/2$ plane of the 3D structure of spin distribution.
The spin textures of other $q_z$ planes with $\delta=\delta_0$ are equivalent to the 2D $q_z$-integrated spin texture with $\delta=\delta_0+\delta'$.
Therefore, by scanning $\delta'$, we can achieve a series of 2D spin textures to reconstruct the 3D spin distribution of given $\delta_0$. This is called Virtual Slicing Image~\cite{Gyu-Boong_Jo,Theory_Liu} (For details, see Methods).
Owing to our stable bias magnetic field~\cite{Magnetic}, $\delta'$ can be tuned precisely.
Fig.\ref{fig2}(b) shows the typical observed 2D spin textures at $\delta_0=-0.5E_{\rm{r}}$, where $q_{z}=0, 0.2\pi, 0.65\pi$ are equivalent to $\delta'=0.396E_{r},0.315E_r,0.176E_r$, respectively.
The spin texture for $q_{z}=0, 0.2\pi$ exhibit band inversion ring patterns~\cite{LZhang2018} corresponding to spin polarizations $P(\boldsymbol{q})=0$,~\cite{Theory_Liu, Wu_Science, Sun_2018A} marked by black circles.
Chern number for each of these 2D bands with fixed $q_{z}$ is determined by the product $\Theta$ of signs of spin polarizations $P(\Lambda_j,q_z)$ at four high symmetric momenta $\left\{\Lambda_{j}\right\}=\left\{\Gamma(0,0),X_{1}(0,\pi),X_{2}(\pi,0),M(\pi,\pi)\right\}$ \cite{C4_Liu,Wu_Science}, $\Theta(q_z)=\prod_j\text{sgn}(P(\Lambda_j,q_z))$.
Thereby scanning the $q_{z}$ planes through the whole 3D Brillouin zone, the topology of the 3D Weyl semimetal band can be fully determined.
Moreover, Weyl nodes emerge on the planes where Chern numbers change by integers.

The stack 2D spin textures in Bloch momentum space with $\delta_0=-0.5E_{\rm{r}}$ is shown in Fig.\ref{fig2}(c).
It is obvious that only two Weyl nodes are here, marked as $\oplus$ and $\ominus$. It corresponds to the IWSM regime.
Black dashed lines sketch the virtual slices of the fusiform band inversion surfaces.
When $q_{z}$ close to 0, spin polarizations near $\Gamma$ point $(0,0,q_{z})$ are negative and outside are positive.
Band inversion ring (black circles in Fig.\ref{fig2}(b)) encloses $\Gamma(0,0,q_{z})$, leads to Chern number $\mathcal{C}=1$ on this 2D plane~\cite{LZhang2018,Sun-2018B}.
As $q_{z}$ goes away from 0 plane, the band inversion rings gradually shrink. Eventually, they vanish at Weyl nodes where $\mathcal{C}$ jumps from 1 to 0.
When $q_{z}$ is close to $\pm\pi$, spin polarizations over the whole layer are positive, and $\mathcal{C}=0$.
From the stack of spin textures in Fig.\ref{fig2}(c), $P(\Lambda_j,q_z)$ are extracted and plotted in Fig.\ref{fig2}(d) as functions of $q_{z}$. $\Theta(q_z)$ is positive for $q_{z}\sim [-\pi,-0.3\pi]$ and $[0.3\pi,\pi]$, while negative for $q_{z}\sim[-0.3\pi,0.3\pi]$. Hence, $\mathcal{C}$ jumps from 0 to 1 at $q_{z}=\pm(0.3\pm0.03)\pi$, indicating the positions of two Weyl points labelled by diamonds.
After removing the high-band thermal effects~\cite{Sun_2018A}, we obtain the corrected locations of Weyl points at $\boldsymbol{q^{\text{W}}}=(0,0,\pm(0.54\pm0.02)E_{\rm{r}})$, agree well with numerical calculations (For details, see Methods).

By varying the detuning term in Eq.(\ref{3DSOC}), we reconstruct the 3D Weyl semimetal bands for $\delta_0=0$ with the same protocol.
To precisely characterize the topology, the high-band effects are taking into account for correction.
The 3D spin structure with for original data (left) and high-band correction (right) are presented in Fig.\ref{fig3}(a).
Three typical 2D spin textures are shown in the insets, with band inversion rings marked.
For $q_z=0.7\pi$, the ring encloses the $\Gamma$ point, indicating $\mathcal{C}=1$. For $q_z=0.3\pi$, the ring encloses the $\text{M}$ point, indicating $\mathcal{C}=-1$. For $q_z=0.5\pi$, the ring touches $\text{X}_1$ and $\text{X}_2$ points, indicating $\mathcal{C}$ jumping by 2 across the plane.
The corresponding four Weyl points locate at $\boldsymbol{q^{\text{W}}}=(0,\pi,\pm(0.52\pm0.03)\pi)$ and $(\pi,0,\pm(0.52\pm0.03)\pi)$, which is also confirmed by the $P(\Lambda_j,q_z)$ and $\Theta(q_z)$ calculation, in Fig.\ref{fig3}(b).

\section{Measuring the Weyl nodes with quench dynamics}

The cold atom system enables the quench studies to characterize the topology with high controllability. We then elucidate the Weyl nodes by quench dynamics. Atom clouds with temperature of 200nK are initially prepared in $\mid\uparrow\rangle$. Raman couplings are effectively excluded with initial detuning $\delta_i=-200E_{r}$. Thereupon, $\delta$ is suddenly switched to $-0.5E_{r}$ in 1$\mu\text{s}$.
Hence, the Hamiltonian is quenched from topological trivial region to IWSM region.
CCD-Y is applied to record the time evolution of spin after the quench, with spin-resolved ToF imaging.

Fig.\ref{fig4}(a) presents the dynamic evolution of spin polarizations $P(q_x,q_z,t)$ in the Brillouin zone, from 0 to 2ms (with $q_y$ being integrated out).
We further get $P(q_z,t)$ by integrating out $q_x$ to emphasise the dynamics along $q_z$ direction, as plotted aside.
For typical $q_z$, $P(q_z,t)$ is plotted as function of evolution time $t$ in Fig.\ref{fig4}(b).
When $q_z$ closed to $\pm\pi$, the evolution of $P(q_z,t)$ is dominated by fast $s$-$p$ band oscillations. 
When $q_z$ closed to $0$, for example $q_z=0$ and $\pm0.3\pi$, $P(q_z,t)$ exhibits both fast and slow oscillations.
We fit those evolutions with a double-frequency damped oscillator (For details, see Methods), and extract the low frequency component $f_{\text{low}}$ as functions of $q_z$, as plotted in Fig.\ref{fig4}(c). Around $q_z=-0.64\pi$ and $q_z=+0.62\pi$, $f_{\text{low}}$ reaches kink minima which we explain below.

The critical points $q_z=-0.64\pi$ and $0.62\pi$ are identified as positions of the two Weyl nodes, close to the result obtained from equilibrium approach.
In topological region with $|q_z|<|q_z^w|$, the 2D bands of $\mid\uparrow\rangle$ and $\mid\downarrow\rangle$ are inverted, with band inversion ring in the 2D planes~\cite{LZhang2018}.
The dynamics is dominated by on-resonance Raman-Rabi oscillations between the two inverted bands~\cite{Sun-2018B, Yi}, corresponding to the low frequency oscillations with large amplitude.
In trivial region, the bands of $\mid\uparrow\rangle$ and $\mid\downarrow\rangle$ are separated, therefore the on-resonance Raman-Rabi oscillations vanish.
At the Weyl points, the two bands touch together at a single point, make $f_{\text{low}}$ reaches minima~\cite{Theory_Liu}, while not zero due to background noises.
The numerical results of low frequency components using experimental parameters are also presented in Fig.\ref{fig4}(c) (see Methods for details), well supporting our observation.

\section{Conclusion and Outlook}

We have realized the 3D SO coupling for ultracold atoms and the ideal Weyl semimetal band (IWSM) which consists of only two Weyl cones and is hard to engineer in solid materials. The IWSM band is experimentally identified by resolving the Weyl points, which are detected by virtual slicing imagining technique and further measured in quench dynamics. We also demonstrate the high tunability by engineering the semimetal band with more Weyl points for comparison. The experimental observations are consistent with the numerical results.

The realization of the IWSM opens a broad avenue to explore exotic quantum phenomena with this optimal Weyl semimetal platform. For example, for the chiral anomaly~\cite{Nielsen1983,Vishwanath2014} the indirect signature was studied in condensed matter physics by measuring negative magnetic resistance~\cite{Huang2015}. The clean and direct probe of this phenomenon may be achieved in the IWSM band with high controllability of the ultracold atoms. Another related but more exotic phenomenon is the chiral magnetic effect~\cite{Fukushima2008}, which states that when an energy shift is introduced between the Weyl points by breaking reflection symmetry, a nonzero chiral current may be generated by a magnetic field, even without applying external electric field~\cite{Franz2013}. This effect is highly debating in condensed matter physics, but may be resolved unambiguously based on current IWSM bands and precise control of atom distributions.

The present protocol for 3D SO coupling and IWSM band is generic, and can be immediately applied to fermion systems, in which case, the various correlated phases shall be accessed by tuning strong interactions. In particular the highly-sought-after topological superfluids~\cite{ZhangC2008,Sato2009} could be achieved for the 3D SO coupled Fermi gases, where the mean-field theory captures essential physics~\cite{Chan2017}, with higher reliability than similar attempts for the 1D or 2D SO coupled systems. 

\begin{addendum}
    \item[Online content] Any methods, extended data, supplementary information, acknowledgements, details of author contributions and competing interests are available online.
\end{addendum}

\bibliographystyle{naturemag}
\bibliography{Weyl-3DSOC}

\begin{thebibliography}{10}
\expandafter\ifx\csname url\endcsname\relax
  \def\url#1{\texttt{#1}}\fi
\expandafter\ifx\csname urlprefix\endcsname\relax\def\urlprefix{URL }\fi
\providecommand{\bibinfo}[2]{#2}
\providecommand{\eprint}[2][]{\url{#2}}

\bibitem{hasan2017discovery}
\bibinfo{author}{Hasan, M.~Z.}, \bibinfo{author}{Xu, S.-Y.},
  \bibinfo{author}{Belopolski, I.} \& \bibinfo{author}{Huang, S.-M.}
\newblock \bibinfo{title}{{Discovery of Weyl Fermion Semimetals and Topological
  Fermi Arc States}}.
\newblock \emph{\bibinfo{journal}{Annual Review of Condensed Matter Physics}}
  \textbf{\bibinfo{volume}{8}}, \bibinfo{pages}{289--309}
  (\bibinfo{year}{2017}).

\bibitem{RMP2018}
\bibinfo{author}{Armitage, N.~P.}, \bibinfo{author}{Mele, E.~J.} \&
  \bibinfo{author}{Vishwanath, A.}
\newblock \bibinfo{title}{{Weyl and Dirac semimetals in three-dimensional
  solids}}.
\newblock \emph{\bibinfo{journal}{Reviews of Modern Physics}}
  \textbf{\bibinfo{volume}{90}}, \bibinfo{pages}{015001}
  (\bibinfo{year}{2018}).

\bibitem{Wan2011}
\bibinfo{author}{Wan, X.}, \bibinfo{author}{Turner, A.~M.},
  \bibinfo{author}{Vishwanath, A.} \& \bibinfo{author}{Savrasov, S.~Y.}
\newblock \bibinfo{title}{{Topological semimetal and Fermi-arc surface states
  in the electronic structure of pyrochlore iridates}}.
\newblock \emph{\bibinfo{journal}{Physical Review B}}
  \textbf{\bibinfo{volume}{83}}, \bibinfo{pages}{205101}
  (\bibinfo{year}{2011}).

\bibitem{Burkov2011}
\bibinfo{author}{Burkov, A.~A.} \& \bibinfo{author}{Balents, L.}
\newblock \bibinfo{title}{{Weyl Semimetal in a Topological Insulator
  Multilayer}}.
\newblock \emph{\bibinfo{journal}{Physical Review Letters}}
  \textbf{\bibinfo{volume}{107}}, \bibinfo{pages}{127205}
  (\bibinfo{year}{2011}).

\bibitem{lv2015experimental}
\bibinfo{author}{Lv, B.~Q.} \emph{et~al.}
\newblock \bibinfo{title}{{Experimental Discovery of Weyl Semimetal TaAs}}.
\newblock \emph{\bibinfo{journal}{Physical Review X}}
  \textbf{\bibinfo{volume}{5}}, \bibinfo{pages}{031013} (\bibinfo{year}{2015}).

\bibitem{TaAs}
\bibinfo{author}{Xu, S.~Y.} \emph{et~al.}
\newblock \bibinfo{title}{{Discovery of a Weyl fermion semimetal and
  topological Fermi arcs}}.
\newblock \emph{\bibinfo{journal}{Science}} \textbf{\bibinfo{volume}{349}},
  \bibinfo{pages}{613--617} (\bibinfo{year}{2015}).

\bibitem{Weyl1929}
\bibinfo{author}{Weyl, H.}
\newblock \bibinfo{title}{{GRAVITATION AND THE ELECTRON}}.
\newblock \emph{\bibinfo{journal}{Proceedings of the National Academy of
  Sciences}} \textbf{\bibinfo{volume}{15}}, \bibinfo{pages}{323--334}
  (\bibinfo{year}{1929}).

\bibitem{Theory_Liu}
\bibinfo{author}{Lu, Y.-H.}, \bibinfo{author}{Wang, B.-Z.} \&
  \bibinfo{author}{Liu, X.-J.}
\newblock \bibinfo{title}{{Realizing and detecting the fundamental Weyl
  semimetal phase}} (\bibinfo{year}{2019}).
\newblock \bibinfo{note}{Preprint at http://arxiv.org/abs/1911.07169}.

\bibitem{Gyu-Boong_Jo}
\bibinfo{author}{Song, B.} \emph{et~al.}
\newblock \bibinfo{title}{{Observation of nodal-line semimetal with ultracold
  fermions in an optical lattice}}.
\newblock \emph{\bibinfo{journal}{Nature Physics}}
  \textbf{\bibinfo{volume}{15}}, \bibinfo{pages}{911--916}
  (\bibinfo{year}{2019}).

\bibitem{No_go}
\bibinfo{author}{Nielsen, H.} \& \bibinfo{author}{Ninomiya, M.}
\newblock \bibinfo{title}{{Absence of neutrinos on a lattice}}.
\newblock \emph{\bibinfo{journal}{Nuclear Physics B}}
  \textbf{\bibinfo{volume}{185}}, \bibinfo{pages}{20--40}
  (\bibinfo{year}{1981}).

\bibitem{IWSM1}
\bibinfo{author}{Zhang, D.} \emph{et~al.}
\newblock \bibinfo{title}{Topological axion states in the magnetic insulator
  $\mathrm{MnBi}_{2}\mathrm{Te}_{4}$ with the quantized magnetoelectric
  effect}.
\newblock \emph{\bibinfo{journal}{Physical Review Letters}}
  \textbf{\bibinfo{volume}{122}}, \bibinfo{pages}{206401}
  (\bibinfo{year}{2019}).

\bibitem{Nielsen1983}
\bibinfo{author}{Nielsen, H.} \& \bibinfo{author}{Ninomiya, M.}
\newblock \bibinfo{title}{{The Adler-Bell-Jackiw anomaly and Weyl fermions in a
  crystal}}.
\newblock \emph{\bibinfo{journal}{Physics Letters B}}
  \textbf{\bibinfo{volume}{130}}, \bibinfo{pages}{389--396}
  (\bibinfo{year}{1983}).

\bibitem{Vishwanath2014}
\bibinfo{author}{Parameswaran, S.~A.}, \bibinfo{author}{Grover, T.},
  \bibinfo{author}{Abanin, D.~A.}, \bibinfo{author}{Pesin, D.~A.} \&
  \bibinfo{author}{Vishwanath, A.}
\newblock \bibinfo{title}{{Probing the Chiral Anomaly with Nonlocal Transport
  in Three-Dimensional Topological Semimetals}}.
\newblock \emph{\bibinfo{journal}{Physical Review X}}
  \textbf{\bibinfo{volume}{4}}, \bibinfo{pages}{031035} (\bibinfo{year}{2014}).

\bibitem{Yao2015PRL}
\bibinfo{author}{Jian, S.-K.}, \bibinfo{author}{Jiang, Y.-F.} \&
  \bibinfo{author}{Yao, H.}
\newblock \bibinfo{title}{{Emergent Spacetime Supersymmetry in 3D Weyl
  Semimetals and 2D Dirac Semimetals}}.
\newblock \emph{\bibinfo{journal}{Physical Review Letters}}
  \textbf{\bibinfo{volume}{114}}, \bibinfo{pages}{237001}
  (\bibinfo{year}{2015}).

\bibitem{Chan2017}
\bibinfo{author}{Chan, C.} \& \bibinfo{author}{Liu, X.-J.}
\newblock \bibinfo{title}{{Non-Abelian Majorana Modes Protected by an Emergent
  Second Chern Number}}.
\newblock \emph{\bibinfo{journal}{Physical Review Letters}}
  \textbf{\bibinfo{volume}{118}}, \bibinfo{pages}{207002}
  (\bibinfo{year}{2017}).

\bibitem{Soluyanov2015}
\bibinfo{author}{Soluyanov, A.~A.} \emph{et~al.}
\newblock \bibinfo{title}{{Type-II Weyl semimetals}}.
\newblock \emph{\bibinfo{journal}{Nature}} \textbf{\bibinfo{volume}{527}},
  \bibinfo{pages}{495--498} (\bibinfo{year}{2015}).

\bibitem{LHuang2016}
\bibinfo{author}{Huang, L.} \emph{et~al.}
\newblock \bibinfo{title}{{Spectroscopic evidence for a type II Weyl
  semimetallic state in MoTe2}}.
\newblock \emph{\bibinfo{journal}{Nature Materials}}
  \textbf{\bibinfo{volume}{15}}, \bibinfo{pages}{1155--1160}
  (\bibinfo{year}{2016}).

\bibitem{Bernevig2016}
\bibinfo{author}{Bradlyn, B.} \emph{et~al.}
\newblock \bibinfo{title}{{Beyond Dirac and Weyl fermions: Unconventional
  quasiparticles in conventional crystals}}.
\newblock \emph{\bibinfo{journal}{Science}} \textbf{\bibinfo{volume}{353}},
  \bibinfo{pages}{aaf5037} (\bibinfo{year}{2016}).

\bibitem{Ding2017}
\bibinfo{author}{Lv, B.~Q.} \emph{et~al.}
\newblock \bibinfo{title}{{Observation of three-component fermions in the
  topological semimetal molybdenum phosphide}}.
\newblock \emph{\bibinfo{journal}{Nature}} \textbf{\bibinfo{volume}{546}},
  \bibinfo{pages}{627--631} (\bibinfo{year}{2017}).

\bibitem{Magnetic1}
\bibinfo{author}{Belopolski, I.} \emph{et~al.}
\newblock \bibinfo{title}{{Discovery of topological Weyl fermion lines and
  drumhead surface states in a room temperature magnet}}.
\newblock \emph{\bibinfo{journal}{Science}} \textbf{\bibinfo{volume}{365}},
  \bibinfo{pages}{1278--1281} (\bibinfo{year}{2019}).

\bibitem{Magnetic2}
\bibinfo{author}{Liu, D.~F.} \emph{et~al.}
\newblock \bibinfo{title}{{Magnetic Weyl semimetal phase in a Kagom{\'{e}}
  crystal}}.
\newblock \emph{\bibinfo{journal}{Science}} \textbf{\bibinfo{volume}{365}},
  \bibinfo{pages}{1282--1285} (\bibinfo{year}{2019}).

\bibitem{Morali2019}
\bibinfo{author}{Morali, N.} \emph{et~al.}
\newblock \bibinfo{title}{{Fermi-arc diversity on surface terminations of the
  magnetic Weyl semimetal Co3Sn2S2}}.
\newblock \emph{\bibinfo{journal}{Science}} \textbf{\bibinfo{volume}{365}},
  \bibinfo{pages}{1286--1291} (\bibinfo{year}{2019}).

\bibitem{IWSM3}
\bibinfo{author}{Ma, J.-Z.} \emph{et~al.}
\newblock \bibinfo{title}{{Spin fluctuation induced Weyl semimetal state in the
  paramagnetic phase of EuCd2As2}}.
\newblock \emph{\bibinfo{journal}{Science Advances}}
  \textbf{\bibinfo{volume}{5}}, \bibinfo{pages}{eaaw4718}
  (\bibinfo{year}{2019}).

\bibitem{IWSM4}
\bibinfo{author}{Jo, N.~H.} \emph{et~al.}
\newblock \bibinfo{title}{{Manipulating of magnetism in the topological
  semimetal EuCd2As2}} (\bibinfo{year}{2020}).
\newblock \bibinfo{note}{Preprint at http://arxiv.org/abs/2002.10485}.

\bibitem{Bloch2013PRL}
\bibinfo{author}{Aidelsburger, M.} \emph{et~al.}
\newblock \bibinfo{title}{{Realization of the hofstadter hamiltonian with
  ultracold atoms in optical lattices}}.
\newblock \emph{\bibinfo{journal}{Physical Review Letters}}
  \textbf{\bibinfo{volume}{111}}, \bibinfo{pages}{185301}
  (\bibinfo{year}{2013}).

\bibitem{Miyake2013PRL}
\bibinfo{author}{Miyake, H.}, \bibinfo{author}{Siviloglou, G.~A.},
  \bibinfo{author}{Kennedy, C.~J.}, \bibinfo{author}{Burton, W.~C.} \&
  \bibinfo{author}{Ketterle, W.}
\newblock \bibinfo{title}{{Realizing the Harper Hamiltonian with Laser-Assisted
  Tunneling in Optical Lattices}}.
\newblock \emph{\bibinfo{journal}{Physical Review Letters}}
  \textbf{\bibinfo{volume}{111}}, \bibinfo{pages}{185302}
  (\bibinfo{year}{2013}).

\bibitem{2D_Esslinger}
\bibinfo{author}{Jotzu, G.} \emph{et~al.}
\newblock \bibinfo{title}{{Experimental realization of the topological Haldane
  model with ultracold fermions}}.
\newblock \emph{\bibinfo{journal}{Nature}} \textbf{\bibinfo{volume}{515}},
  \bibinfo{pages}{237--240} (\bibinfo{year}{2014}).

\bibitem{2D_Bloch}
\bibinfo{author}{Aidelsburger, M.} \emph{et~al.}
\newblock \bibinfo{title}{{Measuring the Chern number of Hofstadter bands with
  ultracold bosonic atoms}}.
\newblock \emph{\bibinfo{journal}{Nature Physics}}
  \textbf{\bibinfo{volume}{11}}, \bibinfo{pages}{162--166}
  (\bibinfo{year}{2015}).

\bibitem{Wu_Science}
\bibinfo{author}{Wu, Z.} \emph{et~al.}
\newblock \bibinfo{title}{{Realization of two-dimensional spin-orbit coupling
  for Bose-Einstein condensates}}.
\newblock \emph{\bibinfo{journal}{Science}} \textbf{\bibinfo{volume}{354}},
  \bibinfo{pages}{83--88} (\bibinfo{year}{2016}).

\bibitem{Bloch2018}
\bibinfo{author}{Lohse, M.}, \bibinfo{author}{Schweizer, C.},
  \bibinfo{author}{Price, H.~M.}, \bibinfo{author}{Zilberberg, O.} \&
  \bibinfo{author}{Bloch, I.}
\newblock \bibinfo{title}{{Exploring 4D quantum Hall physics with a 2D
  topological charge pump}}.
\newblock \emph{\bibinfo{journal}{Nature}} \textbf{\bibinfo{volume}{553}},
  \bibinfo{pages}{55--58} (\bibinfo{year}{2018}).

\bibitem{Song:2017uf}
\bibinfo{author}{Song, B.} \emph{et~al.}
\newblock \bibinfo{title}{{Observation of symmetry-protected topological band
  with ultracold fermions}}.
\newblock \emph{\bibinfo{journal}{Science Advances}}
  \textbf{\bibinfo{volume}{4}}, \bibinfo{pages}{eaao4748}
  (\bibinfo{year}{2018}).

\bibitem{Liu2009}
\bibinfo{author}{Liu, X.-J.}, \bibinfo{author}{Borunda, M.~F.},
  \bibinfo{author}{Liu, X.} \& \bibinfo{author}{Sinova, J.}
\newblock \bibinfo{title}{{Effect of Induced Spin-Orbit Coupling for Atoms via
  Laser Fields}}.
\newblock \emph{\bibinfo{journal}{Physical Review Letters}}
  \textbf{\bibinfo{volume}{102}}, \bibinfo{pages}{046402}
  (\bibinfo{year}{2009}).

\bibitem{Lin2011}
\bibinfo{author}{Lin, Y.-J.}, \bibinfo{author}{Jim{\'{e}}nez-Garc{\'{i}}a, K.}
  \& \bibinfo{author}{Spielman, I.~B.}
\newblock \bibinfo{title}{{Spin-orbit-coupled Bose-Einstein condensates}}.
\newblock \emph{\bibinfo{journal}{Nature}} \textbf{\bibinfo{volume}{471}},
  \bibinfo{pages}{83--86} (\bibinfo{year}{2011}).

\bibitem{Zhang2012}
\bibinfo{author}{Zhang, J.~Y.} \emph{et~al.}
\newblock \bibinfo{title}{{Collective dipole oscillations of a spin-orbit
  coupled Bose-Einstein condensate}}.
\newblock \emph{\bibinfo{journal}{Physical Review Letters}}
  \textbf{\bibinfo{volume}{109}}, \bibinfo{pages}{115301}
  (\bibinfo{year}{2012}).

\bibitem{Baozong}
\bibinfo{author}{Wang, B.~Z.} \emph{et~al.}
\newblock \bibinfo{title}{{Dirac-, Rashba-, and Weyl-type spin-orbit couplings:
  Toward experimental realization in ultracold atoms}}.
\newblock \emph{\bibinfo{journal}{Physical Review A}}
  \textbf{\bibinfo{volume}{97}}, \bibinfo{pages}{011605}
  (\bibinfo{year}{2018}).

\bibitem{JZhang2016}
\bibinfo{author}{Huang, L.} \emph{et~al.}
\newblock \bibinfo{title}{{Experimental realization of two-dimensional
  synthetic spin-orbit coupling in ultracold Fermi gases}}.
\newblock \emph{\bibinfo{journal}{Nature Physics}}
  \textbf{\bibinfo{volume}{12}}, \bibinfo{pages}{540--544}
  (\bibinfo{year}{2016}).

\bibitem{Sun_2018A}
\bibinfo{author}{Sun, W.} \emph{et~al.}
\newblock \bibinfo{title}{{Highly Controllable and Robust 2D Spin-Orbit
  Coupling for Quantum Gases}}.
\newblock \emph{\bibinfo{journal}{Physical Review Letters}}
  \textbf{\bibinfo{volume}{121}}, \bibinfo{pages}{150401}
  (\bibinfo{year}{2018}).

\bibitem{Dubcek2015}
\bibinfo{author}{Dub{\v{c}}ek, T.} \emph{et~al.}
\newblock \bibinfo{title}{{Weyl Points in Three-Dimensional Optical Lattices:
  Synthetic Magnetic Monopoles in Momentum Space}}.
\newblock \emph{\bibinfo{journal}{Physical Review Letters}}
  \textbf{\bibinfo{volume}{114}}, \bibinfo{pages}{225301}
  (\bibinfo{year}{2015}).

\bibitem{Xu2015}
\bibinfo{author}{Xu, Y.}, \bibinfo{author}{Zhang, F.} \&
  \bibinfo{author}{Zhang, C.}
\newblock \bibinfo{title}{{Structured Weyl Points in Spin-Orbit Coupled
  Fermionic Superfluids}}.
\newblock \emph{\bibinfo{journal}{Physical Review Letters}}
  \textbf{\bibinfo{volume}{115}}, \bibinfo{pages}{265304}
  (\bibinfo{year}{2015}).

\bibitem{WangPRA2016}
\bibinfo{author}{Wang, Y.-Q.} \& \bibinfo{author}{Liu, X.-J.}
\newblock \bibinfo{title}{{Predicted scaling behavior of Bloch oscillation in
  Weyl semimetals}}.
\newblock \emph{\bibinfo{journal}{Physical Review A}}
  \textbf{\bibinfo{volume}{94}}, \bibinfo{pages}{031603}
  (\bibinfo{year}{2016}).

\bibitem{Magnetic}
\bibinfo{author}{Xu, X.-T.} \emph{et~al.}
\newblock \bibinfo{title}{{Ultra-low noise magnetic field for quantum gases}}.
\newblock \emph{\bibinfo{journal}{Review of Scientific Instruments}}
  \textbf{\bibinfo{volume}{90}}, \bibinfo{pages}{054708}
  (\bibinfo{year}{2019}).

\bibitem{LZhang2018}
\bibinfo{author}{Zhang, L.}, \bibinfo{author}{Zhang, L.}, \bibinfo{author}{Niu,
  S.} \& \bibinfo{author}{Liu, X.-J.}
\newblock \bibinfo{title}{{Dynamical classification of topological quantum
  phases}}.
\newblock \emph{\bibinfo{journal}{Science Bulletin}}
  \textbf{\bibinfo{volume}{63}}, \bibinfo{pages}{1385--1391}
  (\bibinfo{year}{2018}).

\bibitem{C4_Liu}
\bibinfo{author}{Liu, X.~J.}, \bibinfo{author}{Law, K.~T.},
  \bibinfo{author}{Ng, T.~K.} \& \bibinfo{author}{Lee, P.~A.}
\newblock \bibinfo{title}{{Detecting topological phases in cold atoms}}.
\newblock \emph{\bibinfo{journal}{Physical Review Letters}}
  \textbf{\bibinfo{volume}{111}}, \bibinfo{pages}{120402}
  (\bibinfo{year}{2013}).

\bibitem{Sun-2018B}
\bibinfo{author}{Sun, W.} \emph{et~al.}
\newblock \bibinfo{title}{{Uncover Topology by Quantum Quench Dynamics}}.
\newblock \emph{\bibinfo{journal}{Physical Review Letters}}
  \textbf{\bibinfo{volume}{121}}, \bibinfo{pages}{250403}
  (\bibinfo{year}{2018}).

\bibitem{Yi}
\bibinfo{author}{Yi, C.~R.} \emph{et~al.}
\newblock \bibinfo{title}{{Observing Topological Charges and Dynamical
  Bulk-Surface Correspondence with Ultracold Atoms}}.
\newblock \emph{\bibinfo{journal}{Physical Review Letters}}
  \textbf{\bibinfo{volume}{123}}, \bibinfo{pages}{190603}
  (\bibinfo{year}{2019}).

\bibitem{Huang2015}
\bibinfo{author}{Huang, X.} \emph{et~al.}
\newblock \bibinfo{title}{{Observation of the Chiral-Anomaly-Induced Negative
  Magnetoresistance in 3D Weyl Semimetal TaAs}}.
\newblock \emph{\bibinfo{journal}{Physical Review X}}
  \textbf{\bibinfo{volume}{5}}, \bibinfo{pages}{031023} (\bibinfo{year}{2015}).

\bibitem{Fukushima2008}
\bibinfo{author}{Fukushima, K.}, \bibinfo{author}{Kharzeev, D.~E.} \&
  \bibinfo{author}{Warringa, H.~J.}
\newblock \bibinfo{title}{{Chiral magnetic effect}}.
\newblock \emph{\bibinfo{journal}{Physical Review D}}
  \textbf{\bibinfo{volume}{78}}, \bibinfo{pages}{074033}
  (\bibinfo{year}{2008}).

\bibitem{Franz2013}
\bibinfo{author}{Vazifeh, M.~M.} \& \bibinfo{author}{Franz, M.}
\newblock \bibinfo{title}{{Electromagnetic Response of Weyl Semimetals}}.
\newblock \emph{\bibinfo{journal}{Physical Review Letters}}
  \textbf{\bibinfo{volume}{111}}, \bibinfo{pages}{027201}
  (\bibinfo{year}{2013}).

\bibitem{ZhangC2008}
\bibinfo{author}{Zhang, C.}, \bibinfo{author}{Tewari, S.},
  \bibinfo{author}{Lutchyn, R.~M.} \& \bibinfo{author}{{Das Sarma}, S.}
\newblock \bibinfo{title}{${p}_{x}+i{p}_{y}$ superfluid from $s$-wave
  interactions of fermionic cold atoms}.
\newblock \emph{\bibinfo{journal}{Physical Review Letters}}
  \textbf{\bibinfo{volume}{101}}, \bibinfo{pages}{160401}
  (\bibinfo{year}{2008}).

\bibitem{Sato2009}
\bibinfo{author}{Sato, M.}, \bibinfo{author}{Takahashi, Y.} \&
  \bibinfo{author}{Fujimoto, S.}
\newblock \bibinfo{title}{Non-abelian topological order in $s$-wave superfluids
  of ultracold fermionic atoms}.
\newblock \emph{\bibinfo{journal}{Physical Review Letters}}
  \textbf{\bibinfo{volume}{103}}, \bibinfo{pages}{020401}
  (\bibinfo{year}{2009}).

\end{thebibliography}

\clearpage

\begin{figure}[t!]
\begin{center}
\includegraphics[width=1.0\linewidth]{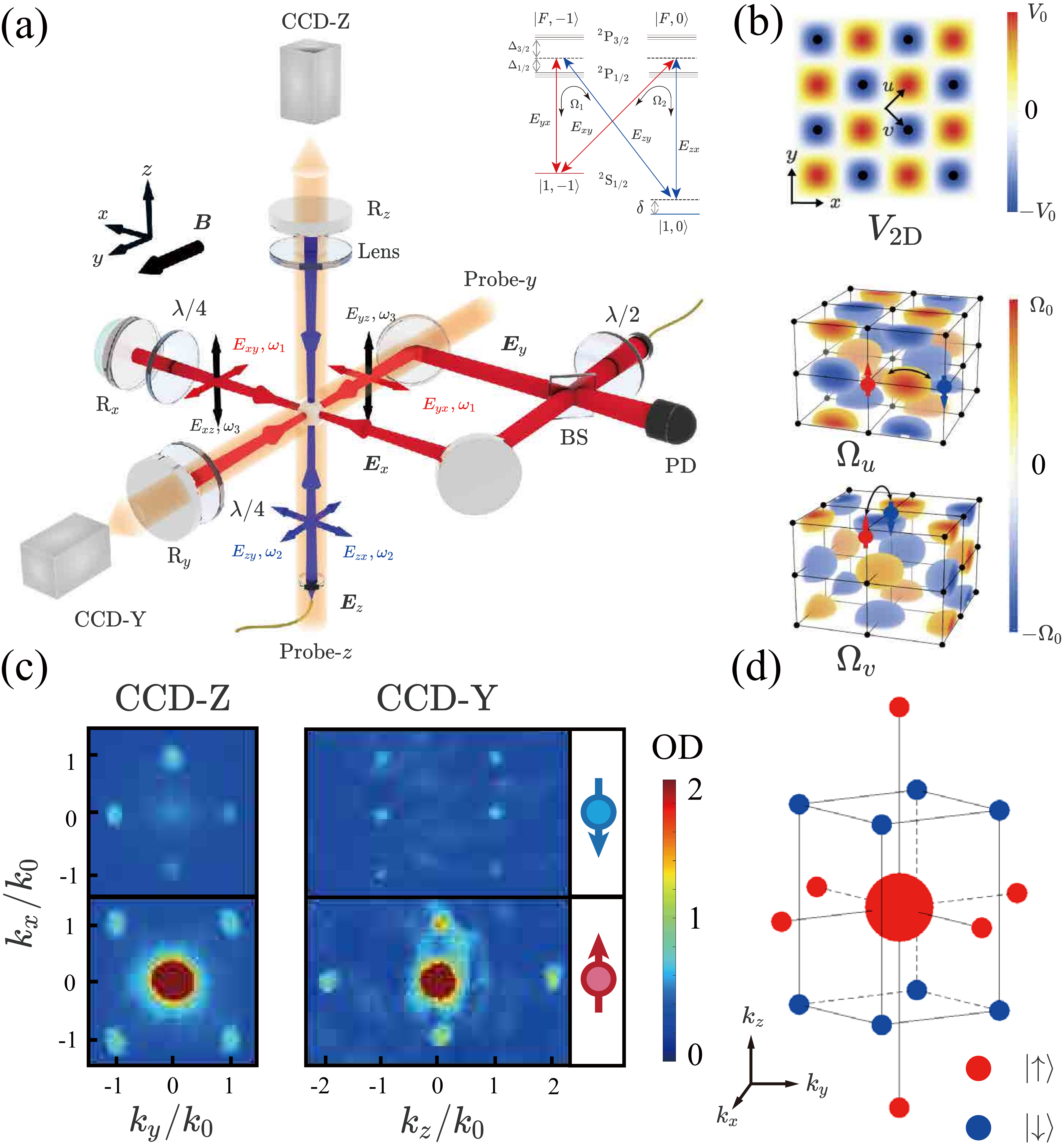}
\vspace{-0.5cm}
\caption{\textbf{Experimental apparatus and 3D SOC.}  %
\label{fig1}}
\end{center}
\end{figure}

\begin{figure}[t!]
\begin{center}
\includegraphics[width=1.0\linewidth]{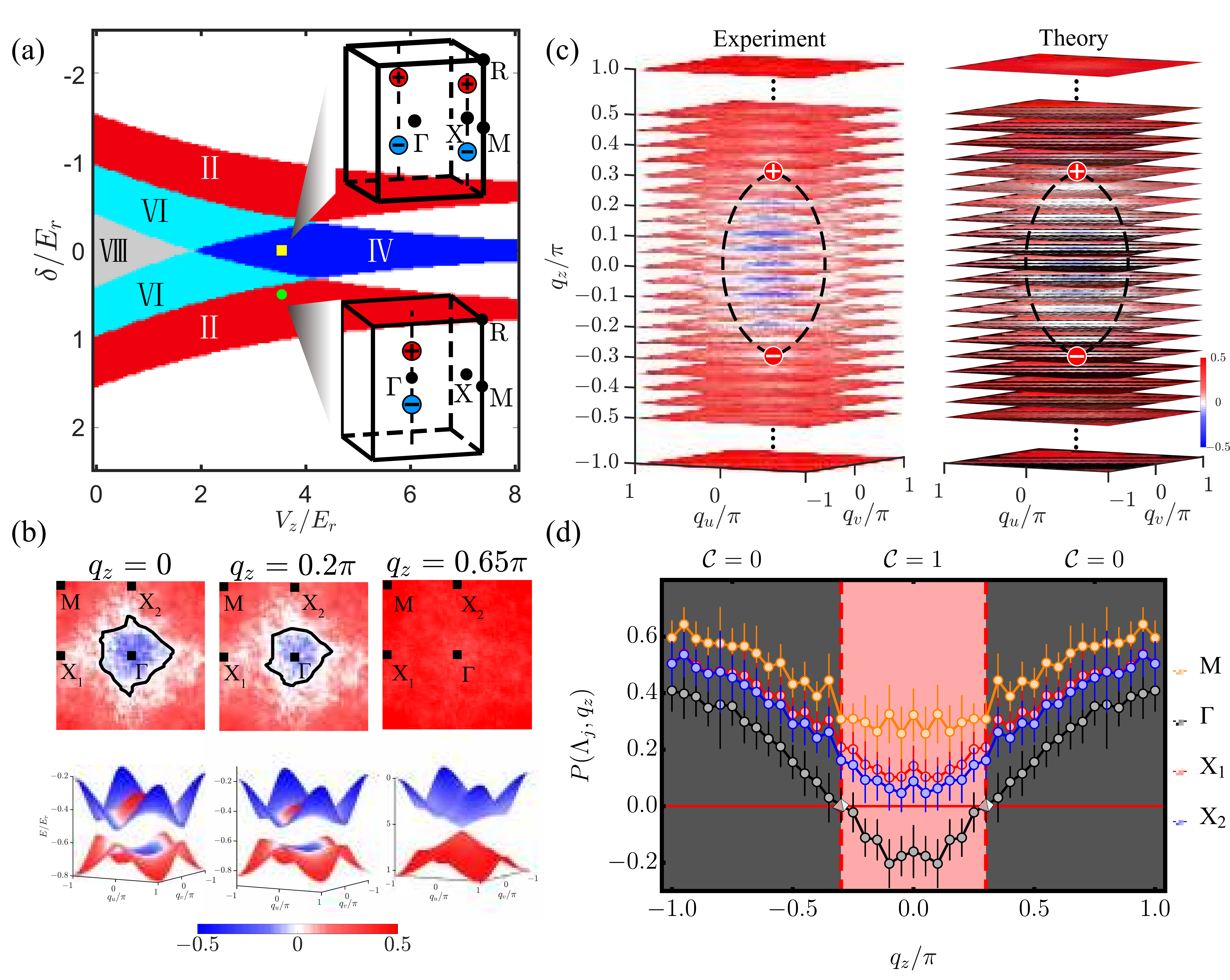}
\vspace{-0.5cm}
\caption{\textbf{Phase diagram and observation of IWSM phase in equilibrium approach.}  %
\label{fig2}}
\end{center}
\end{figure}

\begin{figure}[t!]
\begin{center}
\includegraphics[width=1.0\linewidth]{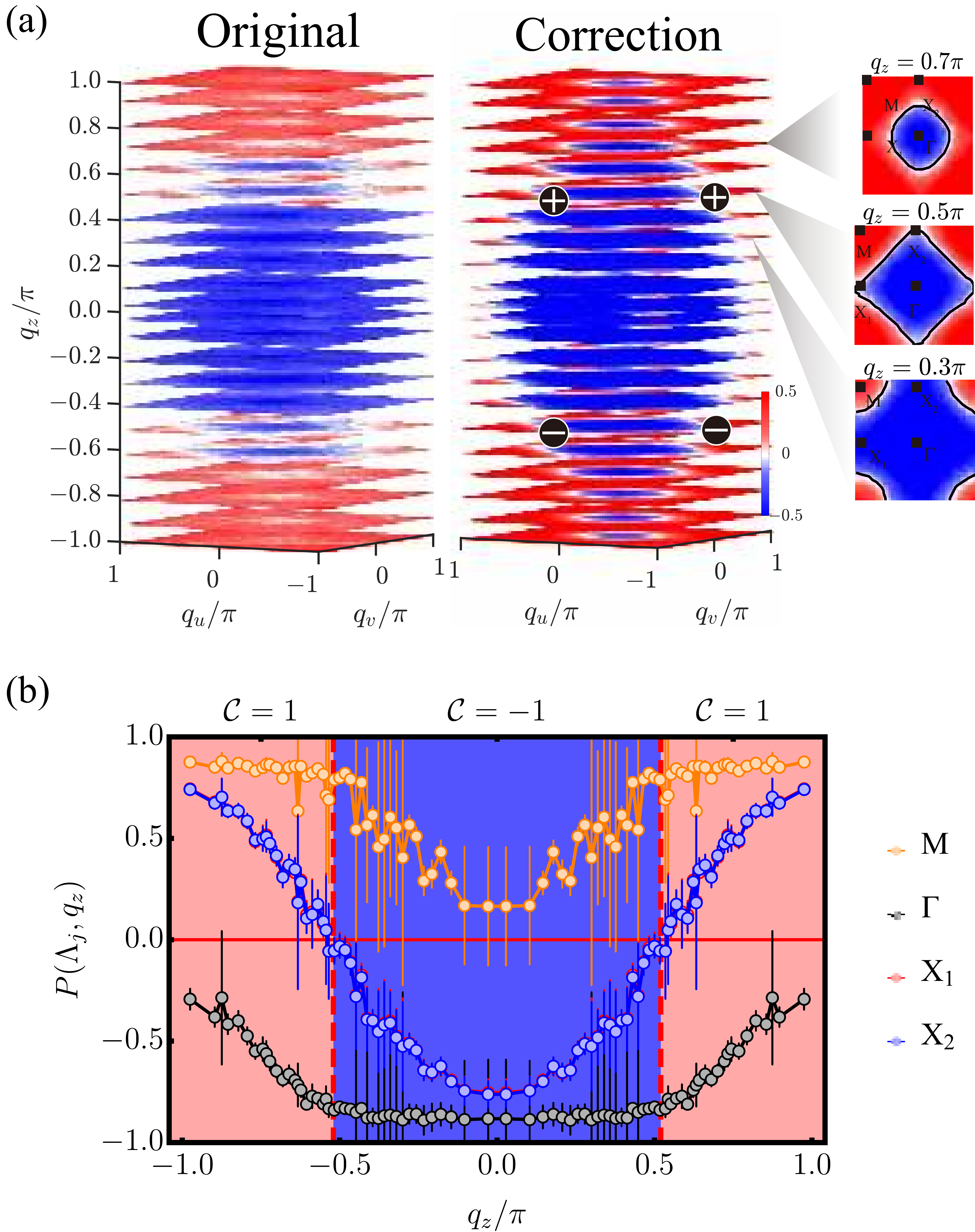}
\vspace{-1.5cm}
\caption{\textbf{Observation of semimetal band with 4 Weyl nodes in equilibrium approach.}  %
\label{fig3}}
\end{center}
\end{figure}

\begin{methods}

\subsection{Realization of 3D Spin-Orbit (SO) coupling and Weyl Hamiltonian.}

As the setup shown in Fig.\ref{fig1}(a) in the main text, 3D SO coupling is realized by three laser beams $\boldsymbol{E}_{i}(i=x,y,z)$ from Ti: sapphire laser, each contains two orthogonal polarization components $E_{ij}(j=x,y,z)$.
$E_{xy}$, $E_{yx}$ have frequency $\omega_{1}$, $E_{xz}$, $E_{yz}$ have frequency $\omega_{2}$, and $E_{zy}$, $E_{zx}$ have frequency $\omega_{3}$. $\boldsymbol{E}_{x}$ and $\boldsymbol{E}_{y}$ form orthogonal standing waves in $x$-$y$ plane. Two $\lambda/4$ waveplates are placed in front of mirror $\boldsymbol{R}_x$ and $\boldsymbol{R}_y$,
producing $\pi$ phase shift between $E_{xz(yz)}$ and $E_{xy(yx)}$. $\boldsymbol{E}_{z}$ forms standing wave along $z$ direction. $\boldsymbol{E}_{\omega_k}(k=1,2,3)$ represent the light fields generated by different frequency components $\omega_k(k=1,2,3)$. $\boldsymbol{E}_{\omega_1}$ is contributed by the polarization component $E_{xy}$ of $\boldsymbol{E}_{x}$ and $E_{yx}$ of $\boldsymbol{E}_{y}$. $\boldsymbol{E}_{\omega_2}$ is contributed by the polarization component $E_{xz}$ of $\boldsymbol{E}_{x}$ and $E_{yz}$ of $\boldsymbol{E}_{y}$. $\boldsymbol{E}_{\omega_3}$ is contributed by the polarization component $E_{zx}$ and $E_{zy}$ of $\boldsymbol{E}_{z}$.  Note that $\omega_1-\omega_2=2\pi\times 200\text{kHz}$, thus it only causes 4mrad phase difference between $\boldsymbol{E}_{\omega_1}$ and $\boldsymbol{E}_{\omega_2}$ over the optical path of 1meter in the experiment, which can be neglected. After eliminating negligible phases and global phases, $\boldsymbol{E}_{\omega_k}(k=1,2,3)$ are written
\begin{equation}\label{1}
\begin{split}
\boldsymbol{E}_{\omega_1}&=\left(\hat{y}e^{\phi_x}E_{xy}\sin{k_{0}x}+
\hat{x}e^{\phi_y}E_{yx}\sin{k_{0}y}\right)e^{-i\omega_1t}\\
\boldsymbol{E}_{\omega_2}&=\hat{z}\left(e^{\phi_x}E_{xz}\cos{k_{0}x}+
e^{\phi_y}E_{yz}\cos{k_{0}y}\right)e^{-i\omega_2t}\\
\boldsymbol{E}_{\omega_3}&=\left(\hat{x}E_{zx}\cos{k_{0}z}+
i\hat{y}E_{zy}\cos{k_{0}z}\right)e^{-i\omega_3t}\\
\end{split}
\end{equation}
where $\phi_{i}(i=x,y)$ are the propagated phases of laser beams from beamsplitter to mirror $\mathbf{R}_i$ along $i$-direction.
Generally, the spin-independent optical potential for typical detuning $\Delta$ is proportional to the light intensity, i.e. $\mathcal{V}_{Latt}\propto\left(\boldsymbol{E}^{*}_{\omega1}\boldsymbol{E}_{\omega1}+\boldsymbol{E}^{*}_{\omega2}\boldsymbol{E}_{\omega2}+\boldsymbol{E}^{*}_{\omega3}\boldsymbol{E}_{\omega3}\right)/\Delta$. 
Hence, the lattice potentials generated by $\omega_k(k=1,2,3)$ read:
\begin{equation}\label{2}
\begin{split}
\mathcal{V}_1&=\frac{1}{3}\left(\frac{\alpha^{2}_{D2}}{\Delta_{3/2}}+\frac{\alpha^{2}_{D1}}{\Delta_{1/2}}\right)\left(\left|E_{xy}\right|^{2}\sin^{2}k_{0}x+\left|E_{yx}\right|^{2}\sin^{2}k_{0}y\right)\\
\mathcal{V}_2&=\frac{1}{3}\left(\frac{\alpha^{2}_{D2}}{\Delta_{3/2}}+\frac{\alpha^{2}_{D1}}{\Delta_{1/2}}\right)\left(\left|E_{xz}\right|^{2}\cos^{2}k_{0}x+\left|E_{yz}\right|^{2}\cos^{2}k_{0}y+2\left|E_{xz}\right|\left|E_{yz}\right|\cos{\phi}\cos{k_{0}x}\cos{k_{0}y}\right)\\
\mathcal{V}_3&=\frac{1}{3}\left(\frac{\alpha^{2}_{D2}}{\Delta_{3/2}}+\frac{\alpha^{2}_{D1}}{\Delta_{1/2}}\right)\left(\left|E_{zx}\right|^{2}+\left|E_{zy}\right|^{2}\right)\cos^{2}k_{0}z\\
\end{split}
\end{equation}
where $\alpha_{D1}$, $\alpha_{D2}$ are the transition dipole matrix elements. $\phi=\phi_x-\phi_y$ is the phase difference between $x$ and $y$ direction. By setting $E_{xy}=E_{xz}$, $E_{yx}=E_{yz}$, and performing a coordinate $45^{\circ}$ rotation ($\hat{u}=(\hat{x}+\hat{y})/\sqrt{2}$, $\hat{v}=(\hat{x}-\hat{y})/\sqrt{2}$), the final lattice potential is obtained in $uvz$-space:
\begin{equation}\label{3}
\begin{split}
\mathcal{V}_{Latt}&=V_{z}\cos^{2}{k_{0}}z+V_{2D}\cos{k_0x}\cos{k_0y}\\
&=V_{z}\cos^{2}{k_{0}}z+V_{2D}\left(\cos^{2}{k_1u}+\cos^{2}{k_1v}\right)
\end{split}
\end{equation}
where $V_{z}=\frac{1}{3}\left(\frac{\alpha^{2}_{D2}}{\Delta_{3/2}}+\frac{\alpha^{2}_{D1}}{\Delta_{1/2}}\right)\left(\left|E_{zx}\right|^{2}+\left|E_{zy}\right|^{2}\right)$, $V_{2D}=\frac{2}{3}\left(\frac{\alpha^{2}_{D2}}{\Delta_{3/2}}+\frac{\alpha^{2}_{D1}}{\Delta_{1/2}}\right)\left|E_{xz}\right|\left|E_{yz}\right|\cos{\phi}$,
$k_{0}=2\pi/\lambda$ and $k_1=k_0/\sqrt{2}$.

The Raman potential $\Omega_R$ is generated by the double-$\Lambda$ scheme from two pairs of polarization components $\left(E_{xy},E_{zx}\right)$, $\left(E_{yx},E_{zy}\right)$ with frequency $\omega_1$ and $\omega_3$. $\omega_2$ does not participate in Raman transitions due to the large detuning ($\delta\approx54E_{r}$), shown in the inset of Fig.1(a) in the main text. $\Omega_R$ reads:
\begin{equation}\label{4}
    \begin{split}
     \Omega_R
     &=\frac{1}{12\sqrt{2}}\left(\frac{\alpha^{2}_{D2}}{\Delta_{3/2}}-\frac{2\alpha^{2}_{D1}}{\Delta_{1/2}}\right)\left(E^{*}_{zy}E_{yx}+E^{*}_{zx}E_{xy}\right)\\
     &=\Omega_{zy}\cos{k_{0}z}\sin{k_{0}y}\sigma_x+e^{i\phi}\Omega_{zx}\cos{k_{0}z}\sin{k_{0}x}\sigma_y
    \end{split}
\end{equation}
where $\Omega_{zy}=\frac{1}{12\sqrt{2}}\left(\frac{\alpha^{2}_{D2}}{\Delta_{3/2}}-\frac{2\alpha^{2}_{D1}}{\Delta_{1/2}}\right)\left|E_{zy}\right|\left|E_{yx}\right|$, $\Omega_{zx}=\frac{1}{12\sqrt{2}}\left(\frac{\alpha^{2}_{D2}}{\Delta_{3/2}}-\frac{2\alpha^{2}_{D1}}{\Delta_{1/2}}\right)\left|E_{zx}\right|\left|E_{xy}\right|$.
By setting $\Omega_{zx}=\Omega_{zy}=\Omega_{0}$ and $\phi=0$, $\Omega_R$ in $uvz$-space is
\begin{equation}\label{5}
    \begin{split}
     \Omega_R&=\sqrt{2}\Omega_{0}\cos{k_{0}z}(\sin{k_1u}\cos{k_1v}\sigma_x+\cos{k_1u}\sin{k_1v}\sigma_y)
    \end{split}
\end{equation}

From Eq.(\ref{3}) and Eq.(\ref{4}), one finds that $\phi$ not only modulates the lattice depth $\mathcal{V}_{2D}$ in $x$-$y$ plane but also adjusts the relative phase between two Raman processes. $\phi=0$ is essential to build 2D chequerboard lattices $\mathcal{V}_{Latt}$ and realise 3D SOC. In the experiment, we lock $\phi$ to 0 (see next section). Combining $\mathcal{V}_{Latt}$ and $\Omega_{R}$, we eventually demonstrate the 3D SOC Hamiltonian in Eq.(2) in the main text.

To have a clear picture of 3D SOC Hamiltonian, we turn to the Bloch momentum space. Under the tight-binding (TB) approximation, the spin-reserved (spin-flipping) hopping terms are determined by the overlap integral of Wannier wavefunctions (plus Raman potential), respectively. Raman couplings induced hopping terms read
\begin{equation}\label{5}
t^{j,j\pm1_{(u)}}_{\text{SO}}=\pm(-1)^{j_u+j_v+j_z}t_{\text{SO}}, \ t^{j,j\pm1_{(v)}}_{\text{SO}}=\pm(-1)^{j_u+j_v+j_z}t_{\text{SO}}
\end{equation}
where $t_{\text{SO}}$ is the amplitude. To remove the staggered sign, we perform a gauge transformation $U=e^{(k_1u+k_1v+k_0z)\mid\uparrow\rangle\langle\uparrow\mid}$ such that Weyl type Hamiltonian are obtained as Eq.(3) in the main text.

\subsection{Phase lock of $\phi$.}


To lock phase $\phi=0$, a Michaelson interferometer in $x$-$y$ plane is applied. The phase-lock setup is shown in Extended Data Fig.1(a). In the experiment, another laser beam with wavelength of 767nm is utilizeed to lock $\phi$.
A negative feedback loop is established to lock the relative optical path of the Michaelson interferometer. The phase noise is evaluated by monitoring the signal of the 767nm-laser from a photodiode (PD1). Its value is below $0.4^\circ$ with different reference voltage after lock.

Another photodiode (PD2) is used to monitor the phase of lattice beams when the Michaelson interferometer is locked by 767nm-laser. Due to the length difference between two branches of the interferometer, $\phi$ can be adjusted by slightly tuning the wavelength of lattice beams. The phase $\phi_{PD}$ detected by PD2 has a relationship with $\phi$: $\phi_{PD}=2\phi+\pi$.
In the experiment, we set $\phi_{PD}=\pi$, thus $\phi=0$. The noise of $\phi$ here is well below $3^\circ$, shown in Extended Data Fig.1(b), satisfying the experimental requirement. Further, we adiabatically load the chequerboard lattices with different $\phi_{PD}$ to test the behaviour of the phase lock. In Extended Data Fig.1(c), we measure the fraction of the lattice diffraction of the total atom numbers, indicating the change of lattice depth. As can be seen, the lattice depth is modulated by $\phi_{PD}$ and it reaches the maximum at $\phi_{PD}=\pi$.

\subsection{ Reconstruction of the first Brillouin zone.}
In the experiment, for detection, spin-resolved time-of-flight (ToF) absorption imaging is performed. Whereas the ToF image displays the bare momentum distribution, we need to reconstruct it in the first Brillouin zone (FBZ) to characterise topology of the Hamiltonian. The eigenstates of the Hamiltonian can be expanded with Bloch functions:
\begin{equation}
    \Psi=\Psi_{\uparrow}+\Psi_{\downarrow}=\sum_{mnl}a_{mnl}\psi_{mnl}^{\uparrow}\chi^{\uparrow}+\sum_{pqr}a_{pqr}\psi_{pqr}^{\downarrow}\chi^{\downarrow},
\end{equation}
where
\begin{equation}
    \begin{split}
        \psi_{mnl}^{\uparrow}&=\text{e}^{i(q_u+2mk_1+k_1)u+i(q_v+2nk_1+k_1)v+i(q_z+2lk_0+k_0)z}\\ \nonumber
        \psi_{pqr}^{\downarrow}&=\text{e}^{i(q_u+2pk_1)u+i(q_v+2qk_1)v+i(q_z+2rk_0)z}
    \end{split}
\end{equation}
and \(\chi^{\uparrow}\)(\(\chi^{\downarrow}\)) represents the $\mid\uparrow\rangle$ ($\mid\downarrow\rangle$) state of the atom. One can see that the eigenstate at each quasimomentum \((q_u,q_v,q_z)\) is the superposition of real momentum $(q_u+2mk_1+k_1,q_v+2nk_1+k_1,q_z+2lk_0+k_0)$ for $\mid\uparrow\rangle$ and $(q_u+2mk_1,q_v+2nk_1,q_z+2lk_0)$ for $\mid\downarrow\rangle$. The extra momentum shift for $\mid\uparrow\rangle$ comes from Raman coupling, transferring \((k_1,k_1,k_0)\) momentum to $\mid\uparrow\rangle$ in quasimomentum space. The size of the FBZ can be determined from the ground state figures, denoted by red rectangles, shown in Extended Data Fig.\ref{FBZ}(a) and (c).

To obtain spin texture in the FBZ, we need to transform the real momentum space into quasimomentum space. For CCD-Z imaging, it is convenient to reconstruct the FBZ under \(q_u,q_v\) axes, where the primitive cells can be chosen as squares, shown in Extended Data Fig.\ref{FBZ}(a) and (b). The real momentum origin of $\mid\uparrow\rangle$ and $\mid\downarrow\rangle$ correspond to quasimomentum \((k_1,k_1)\) and \((0,0)\). For CCD-Y imaging, the primitive cells are rectangles, shown in Extended Data Fig.\ref{FBZ}(c) and (d). The real momentum origin of $\mid\uparrow\rangle$ and $\mid\downarrow\rangle$ correspond to quasimomentum \((0,k_0)\) and \((0,0)\). The reconstruction procedure is as follows: First, we determine the size of FBZ from the ground state as the size of the primitive cells. Second, we divide the photos of thermal atom cloud into these primitive cells.
 After that, we add the atom distribution \(N^{\uparrow}_{FBZ}(\boldsymbol{q})\) and \(N^{\downarrow}_{FBZ}(\boldsymbol{q})\) in the primitive cells together. Finally, we calculate the spin texture \(P(\boldsymbol{q})=(N^{\uparrow}_{FBZ}(\boldsymbol{q})-N^{\downarrow}_{FBZ}(\boldsymbol{q}))/(N^{\uparrow}_{FBZ}(\boldsymbol{q})+N^{\downarrow}_{FBZ}(\boldsymbol{q}))\).

\subsection{Virtual slicing image: \(\delta^{\prime}\)-\(q_z\) mapping. }
Detecting 3D bands from a 2D image is a challenge. Fortunately, the emergent magnetic group symmetry of our system helps us to reconstruct the 3D bands from a series of 2D spin textures via virtual slicing approach \cite{Gyu-Boong_Jo,Theory_Liu}.
In this approach, the measured 2D spin textures for different $\delta$ is consistent with the total spin texture acquired numerically with the same $\delta$ based on the 3D Hamiltonian Eq.(\ref{3DSOC}) (for a comparison of experimental with theoretical results, see Extended Data Fig.\ref{Virtual}). This measurement confirms the realization of the expected 3D Weyl semimetal bands.

The mapping between \(\delta'\) and \(q_z\) can be analytically proved under the Tight-Binding (TB) approximation. The Hamiltonian in Bloch momentum space reads:
\begin{equation}
    \begin{split}
        H_{\text{TB}}
        &= [\frac{\delta}{2} - 2t_z\cos q_z- 2t_0(\cos q_u + \cos q_v) ]\sigma_z
        + 2t_\text{SO}\sin q_u\sigma_x + 2t_\text{SO}\sin q_v\sigma_y
    \end{split}
\end{equation}
And the 3D spin texture of the lowest band is $P(\boldsymbol{q}) = \langle\sigma_z\rangle(\boldsymbol{q})$. For term $\delta/2-2t_z\cos q_z$, it's easy to vary \(\delta/2\) and \(q_z\) together, but maintain that term unchanged, thus the Hamiltonian of the system is invariant. Therefore, to observe the 2D spin texture of different \(q_z\) planes in the system of detuning \(\delta_0\)
of interest, one shall instead plot 2D spin textures of a fixed \(q_{z0}\) plane with different \(\delta\), the correspondence equation being
\begin{equation}
    \delta'= -4t_z(\cos q_z-\cos q_{z0})
\end{equation}
Under such condition, we have spin texture
\begin{equation}
    \label{eq5}
    P(q_u,q_v,q_{z0};\delta)=P(q_u,q_v,q_{z};\delta_0).
\end{equation}

Given the emergent magnetic group symmetry, spin texture is anti-symmetric with respect to $q_z=\pi/2$ along $q_z$ direction. Hence, one can prove that the 2D spin texture of \(q_z=\pi/2\) plane in the system of \(\delta\) has the same sign as the integrated 2D spin texture of this system. i.e.
\begin{equation}
    \label{eq6}
    \mathrm{sgn}[\frac{1}{2\pi} \int^{2\pi}_{0} P(q_u,q_v,q_z;\delta)
    \mathrm dq_z] = \mathrm{sgn}[P(q_u,q_v,q_{z0} = \pi/2;\delta)],
\end{equation}
Thus, combining Eq.(\ref{eq5}) and Eq.(\ref{eq6}), one can obtain
\begin{equation}\label{eq7}
    \mathrm{sgn}[\frac{1}{2\pi} \int^{2\pi}_{0} P(q_u,q_v,q_z;\delta) \mathrm dq_z] =
    \mathrm{sgn}[P(q_u,q_v,q_{z};\delta_0)].
\end{equation}
Hence, one can obtain 2D spin textures of different \(q_z\) of \(\delta_0\) by measuring the integrated 2D spin textures of different \(\delta=\delta_0+\delta'\).

For the real system, the sign equivalence in Eq.(\ref{eq7}) still holds.
The correspondence between $\delta^{\prime}$ and $q_z$ can be numerically verified with plane wave expansion. Correspondences for different \(\delta_0\) in the experiment are calculated in Extended Data Table.1 and Table.2.

\subsection{High-band correction.}
To identify band topology by $\Theta(q_z)$ in the main text, it requires that atoms occupy only at the lowest band. Usually, as temperature rises, more atoms populate at high bands, and calculating the spin polarization should take the atoms of high bands into account. Those thermal atoms change the value of spin polarizations of the lowest band, and decrease the signal to noise ratio of the net spin polarization. This affects crucially the accuracy of distinguishing topology, and we have seen it in Ref \cite{Sun_2018A}.

We numerically eliminate the high-band thermal effects by subtracting atoms occupying at high bands from the total atom density. The corrected spin polarization is:
\begin{equation}
    P_s(\boldsymbol{q})=\frac{n_\uparrow(\boldsymbol{q};T)\gamma_\uparrow(\boldsymbol{q};T)-n_\downarrow(\boldsymbol{q};T)\gamma_\downarrow(\boldsymbol{q};T)}{n_\uparrow(\boldsymbol{q};T)\gamma_\uparrow(\boldsymbol{q};T)+n_\downarrow(\boldsymbol{q};T)\gamma_\downarrow(\boldsymbol{q};T)}
\end{equation}
where \(\gamma_\sigma(\boldsymbol{q};T)=n_{\sigma,s}(\boldsymbol{q};T)/n_\sigma(\boldsymbol{q};T)\) (\(\sigma=\uparrow,\downarrow\)) is the fraction in the lowest band with Bose distribution, and $T$ is temperature.
Extended Data Fig.\ref{RemoveHB} shows the corrected 2D spin texture stack and spin polarizations of high symmetric points around \(T=150\)nK. The contrast of the figures as well as the signal to noise ratio of spin polarizations are significantly improved. The Weyl points for \(\delta_0=-0.5E_r\) locate at \(q_z=\pm(0.54\pm0.03)\pi\), agreeing quite well with the numerical result \(q_z=\pm0.6\pi\).

\subsection{Fitting of dynamic evolution. }
To extract topological information from quench dynamics, we employ a model of double-frequency damped oscillator to fit the time evolution of spin polarizations on each \(q_z\), i.e.
\begin{equation}
    P_{\text{fit}}(t)\mid_{q_z}=a_1\mathrm{e}^{-t^2/\tau_1^2}\cos(2\pi f_1t+\varphi_1)+a_2\mathrm{e}^{-t^2/\tau_2^2}\cos(2\pi f_2t+\varphi_2)+a_0\mathrm{e}^{-t/\tau_0}+s_0
\end{equation}
where $a_1$, $a_2$, $a_0$, $\tau_1$, $\tau_2$, $\tau_0$, $f_1$, $f_2$, $\varphi_1$, $\varphi_2$, $s_0$ are fitting parameters. The lower frequency $\min[f_1,f_2]$ represents oscillations between two s-bands, while the higher frequency $\max[f_1,f_2]$ represents oscillations between s-bands and higher bands. Each frequency term has an exponential decay, with decay time \(\tau_1\) and \(\tau_2\). $\tau_0$ term is the constant decay of the background. This model is totally empirical based on the behaviour of experimental data.
As examples, we fit two typical oscillations of spin polarization.

For \(q_z=0.214\pi\), the 2D spin texture lies within the topological zone with the s-bands touching and creating a band inversion ring (Extended Data Fig.\ref{QuenchExp}(a)). At the ring, spin polarization of the post-quench strongly oscillates between the s bands. Whereas away from the ring, spin polarization oscillates weakly.
After integrating out \(q_x\) and \(q_y\), coarsening along \(q_z\) direction, the oscillation (Extended Data Fig.\ref{QuenchExp}(c)) reflects the on-resonance Raman-Rabi oscillation. One can see that the oscillation consists of two major frequency components, a low frequency oscillation that lasts for about 1ms and a high frequency one that quickly decays around 200\(\mu\text{s}\). Meanwhile, the steady value of the oscillation declines over the whole evolution, signifying an exponential background decay. Fit gives \(f_{\text{low}}=1.08\text{kHz}\times(1\pm0.032)\), \(f_{\text{high}}=5.10\text{kHz}\times(1\pm0.11)\), \(\tau_{\text{low}}=1.49\text{ms}\times(1\pm0.14)\) and \(\tau_{\text{high}}=0.072\text{ms}\times(1\pm0.098)\).

For \(q_z=0.990\pi\), the 2D spin texture is trivial with s-bands far away separated (Extended Data Fig.\ref{QuenchExp}(b)). The on-resonance Raman-Rabi oscillations vanish. The oscillation (Extended Data Fig.\ref{QuenchExp}(d)) is dominated by high frequency component. Fit gives \(f_{\text{low}}=0.68\text{kHz}\times(1\pm0.064)\), \(f_{\text{high}}=9.31\text{kHz}\times(1\pm0.045)\), \(\tau_{\text{low}}=1.39\text{ms}\times(1\pm0.18)\) and \(\tau_{\text{high}}=0.15\text{ms}\times(1\pm0.17)\).


\subsection{Numerical simulation of quench dynamics. }
To verify the locations of Weyl nodes identified in the quench dynamics, we numerically demonstrate the connection between oscillation modes and the positions of Weyl nodes. For convenience, we derive an effective tight-binding (TB) model by including $s$, $p_x$, and $p_y$ bands, described by
\begin{equation*}
h\left(\boldsymbol{q}\right)=\left(\begin{array}{ccc}
h_s\left(\boldsymbol{q}\right) & t_{\rm so}^{s,p_x}\sigma_{x} & t_{\rm so}^{s,p_y}\sigma_{y}\\
t_{\rm so}^{s,p_x} \sigma_{x} & h_{p_x}\left(\boldsymbol{q}\right) & t_{\rm so}^{p_x,p_y}\left(\sigma_{y}-\sigma_{x}\right)\\
t_{\rm so}^{s,p_y}\sigma_{y} & t_{\rm so}^{p_x,p_y}\left(\sigma_{y}-\sigma_{x}\right) & h_{p_y}\left(\boldsymbol{q}\right)
\end{array}\right).
\end{equation*}
Here  $h_{s}(\boldsymbol{q})$, $h_{p_x}(\boldsymbol{q})$ and $h_{p_y}(\boldsymbol{q})$ describe  $s$, $p_x$ and $p_y$ bands, respectively
\begin{eqnarray}\label{eqTB}
    h_s\left(\boldsymbol{q}\right) &=& \left[m_{z}-2t_{z}\cos{q_{z}}-2t_{z2}\cos{2q_{z}}-2t_{1}\left(\cos{q_{u}} +\cos{q_{v}}\right)-2t_{2}\left(\cos{2q_{u}}+\cos{2q_{v}}\right)\right]\sigma_{z} \nonumber\\
     && \quad+2t_\text{so}(\sin{q_{u}}\sigma_{x}+\sin{q_{v}}\sigma_{y})  \\
    h_{p_x}\left(\boldsymbol{q}\right) &=& \epsilon_{\rm sp}+\left(m_{z}+2t_{p}\cos q_{u}+2t_{p2}\cos 2q_{u}\right) \sigma_{z}+2t_{\rm so}^p \sin q_{v} \sigma_{y},\nonumber\\
    h_{p_y}\left(\boldsymbol{q}\right) &=& \epsilon_{\rm sp}+\left(m_{z}+2t_p \cos q_{v} +2t_{p2} \cos 2q_{v}\right) \sigma_{z}-2t_{\rm so}^{p} \sin q_{u} \sigma_{x}.\nonumber
\end{eqnarray}
where $\epsilon_{\rm sp}$ is the difference of self energy between $s$ and $p$ orbitals. The hopping coefficients are derived from experimental conditions with the help of Wannier functions, and fine-tuned to match the exact diagonalization of the Hamiltonian in continuous space. For this the next-nearest-neighbor hopping is also considered. All the parameters in the TB model are showed in Extended Data Table.3.

Then the calculation of quench dynamics is performed with Eq.\ref{eqTB}. The initial state is fully polarized $\mid\uparrow \rangle$, and the post-quench Hamiltonian has two Weyl nodes according to the experimental condition. During the time evolution, the system inevitably interact with environment and the amplitude of oscillation decays with time. This process can be captured by Lindblad Master equation, describing the evolution of density matrix. We found that the numerical results match the experimental observations well when taking the damping factor to properly depend on the spin polarization and energy difference of post-quench Hamiltonian. In particular, the damping factor is set small near the band inversion rings and energy gap, whereas it increases rapidly away from the such regions. In this case, the oscillation of spin polarization is dominated by the states around band inversion rings and energy gap, implying the actual spin dynamics measured in the experiment. Integrating out the oscillations in the $q_x$-$q_y$ plane, we obtain the oscillations for different $q_z$. The numerical results are shown in Extended Data Fig.\ref{Numeric}, matching experimental data in Fig.\ref{fig4}. The evolution of $P(q_z,t)$ is dominated by fast oscillations in trivial region while it exhibits both fast and slow behaviours in topological region. Finally we fit the frequency and obtain $f_{low}$ VS $q_z$ in Fig.\ref{fig4}(c) as comparison.
The minima of frequency curve denote the positions of Weyl nodes.

\end{methods}

\begin{addendum}
  \item We acknowledge insightful discussions with Long Zhang and thanks Wei Sun and Xiao-Tian Xu for their experimental preparation in early stage.
  This work was supported by the National Key R\&D Program of China (under grants 2016YFA0301601 and 2016YFA0301604), National Natural Science Foundation of China (under grants No. 11674301, 11825401, 11761161003, and 11921005), the Anhui Initiative in Quantum Information Technologies (AHY120000), and the Strategic Priority Research Program of Chinese Academy of Science (Grant No. XDB28000000).
  \item[Author contributions] Put author contributions here.
  \item[Competing interests] The authors declare that they have no competing financial interests.
  \item[Correspondence] Correspondence and requests for materials should be addressed to X.J.L. (xiongjunliu@pku.edu.cn);
S.C. (shuai@ustc.edu.cn); or J.W.P. (pan@ustc.edu.cn).
  \item[Data availability] The data that support the plots within this paper and other findings of this study are available from the corresponding author upon reasonable request.
  \item[Code availability] The code that support the plots within this paper are available from the corresponding author upon reasonable request.

\end{addendum}

\clearpage

\textbf{Figure 1: Experimental apparatus and 3D SOC.}
\textbf{a.} Experimental setup and the double-$\Lambda$ configuration of Raman coupling. Laser beams $\boldsymbol{E}_{x}$, $\boldsymbol{E}_{y}$, and $\boldsymbol{E}_{z}$ are from Ti: sapphire laser, producing 3D optical lattices with Raman couplings.
\textbf{b.} Density plot of chequerboard lattice potential in real space (top). The antisymmetric structure of the two Raman couplings $\Omega_{u}$ (middle) and $\Omega_{v}$ (bottom). The grids represent lattice sites of $\mathcal{V}_{Latt}$.
\textbf{c.} Spin-resolved ToF images of ground state of 3D SO coupled BEC along $\hat{z}$ and $\hat{y}$, respectively, where $V_{2D}=1.77E_{r}$, $V_{z}=3.54E_{r}$, $\Omega_{0}=1.02E_{r}$ and $\delta=-0.1E_{r}$.
\textbf{d.} Illustration of reconstruction of momentum distribution.

\textbf{Figure 2: Phase diagram and observation of IWSM phase in equilibrium approach.}
\textbf{a.} Phase diagram of Weyl semimetal band with $V_{2D}=1.77E_{r}$ and $\Omega_{0}=1.02E_{r}$. Numbers of Weyl points are written by Roman numerals and distinguished by different colors. Green circle and yellow square are corresponding to experimental points. The insets are 3D Brillouin for 4 Weyl points (upper) and IWSM (lower). Weyl points are labeled by $\oplus$ or $\ominus$. The four high symmetric points are marked by $\Gamma$(0,0,0), M($\pi$,$\pi$,0), X(0,$\pi$,0), R($\pi$,$\pi$,$\pi$).
\textbf{b.} Three typical spin textures, imaging on CCD-Z. In the experiment, $V_{2D}=1.77E_{r}$, $V_{z}=3.54E_{r}$, and $\Omega_{0}=1.02E_{r}$. The first Brillouin zone is shifted with $\Gamma(0,0)$ in the center. Four high symmetric momenta $(\text{M},\Gamma,\text{X}_{1},\text{X}_{2})$ are marked in the spin textures. Black circles point out the band inversion rings. The corresponding band structures are shown below.
\textbf{c.} Reconstruction of spin textures of different $q_{z}$ layers in 3D Bloch momentum space with $\delta_0=-0.5E_{r}$. The left (right) stack is experimental data (numerical calculation). Weyl points are marked by $\oplus$ or $\ominus$. Two black dashed lines depict the fusiform of band inversion surface.
\textbf{d.} Spin polarizations $P(\Lambda_j,q_z)$ at four high symmetric momenta versus $q_{z}$.  Weyl points are marked by diamonds.

\textbf{Figure 3: Observation of semimetal band with 4 Weyl nodes in equilibrium approach.}
\textbf{a.} Reconstruction of spin textures of different $q_{z}$ layers in 3D Bloch momentum space with $\delta_0=0$. The left (right) stack is original experimental data (high-band correction data). Weyl points are marked by $\oplus$ or $\ominus$. Three 2D spin textures are shown to demonstrate topology change around 4 Weyl points.
\textbf{b.} Spin polarizations $P(\Lambda_j,q_z)$ at four high symmetric momenta versus $q_{z}$.

\textbf{Figure 4: Measuring the Weyl nodes with quench dynamics.}
\textbf{a.} Dynamic evolution of spin polarizations in the Brillouin zone, imaging on CCD-Y. In the experiment, $V_{2D}=1.77E_{r}$, $V_{z}=3.54E_{r}$, $\Omega_{0}=1.02E_{r}$ and $\delta_0=-0.5E_{r}$.
\textbf{b.} Typical oscillations of spin polarization $P(q_z,t)$ for different $q_{z}$. Red circles with error bars are experimental data. Blue curves are fitted by a double-frequency damped oscillator model.
\textbf{c.} The fitting parameters $f_{low}$ versus $q_{z}$ both in experimental data (blue circles with error bars) and numerical calculation (black circles). $\oplus$ and $\ominus$ mark the locations of two Weyl points.

\clearpage

\begin{figure}[t!]
\begin{center}
\includegraphics[width=1.0\linewidth]{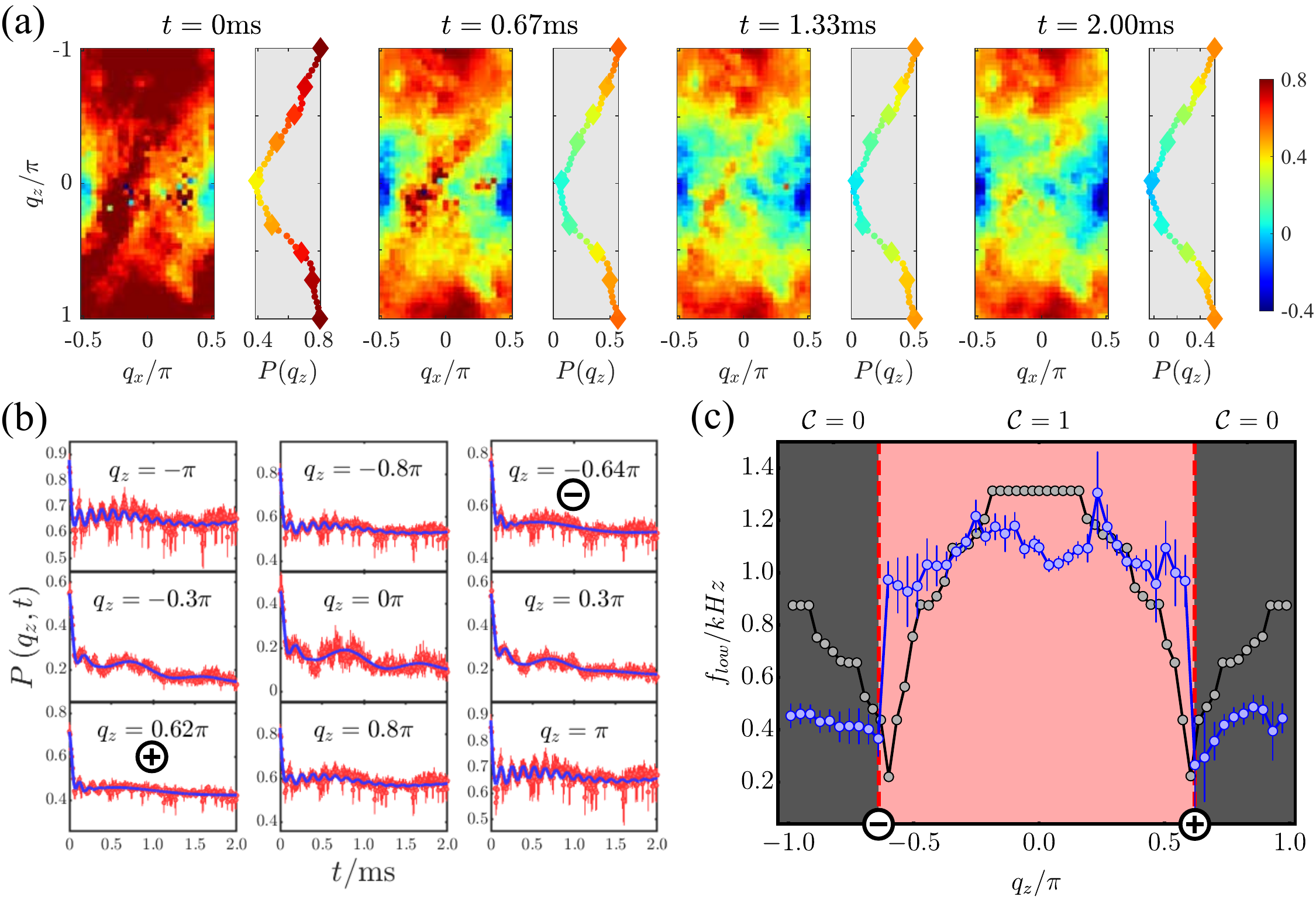}
\vspace{-0.5cm}
\caption{\textbf{Measuring the Weyl nodes with quench dynamics.}  %
\label{fig4}}
\end{center}
\end{figure}

\setcounter{figure}{0}
\renewcommand{\figurename}{Extended Data Figure}
\renewcommand{\thefigure}{\arabic{figure}}

\begin{figure}[h]
    \centering
    \includegraphics[width=1\linewidth]{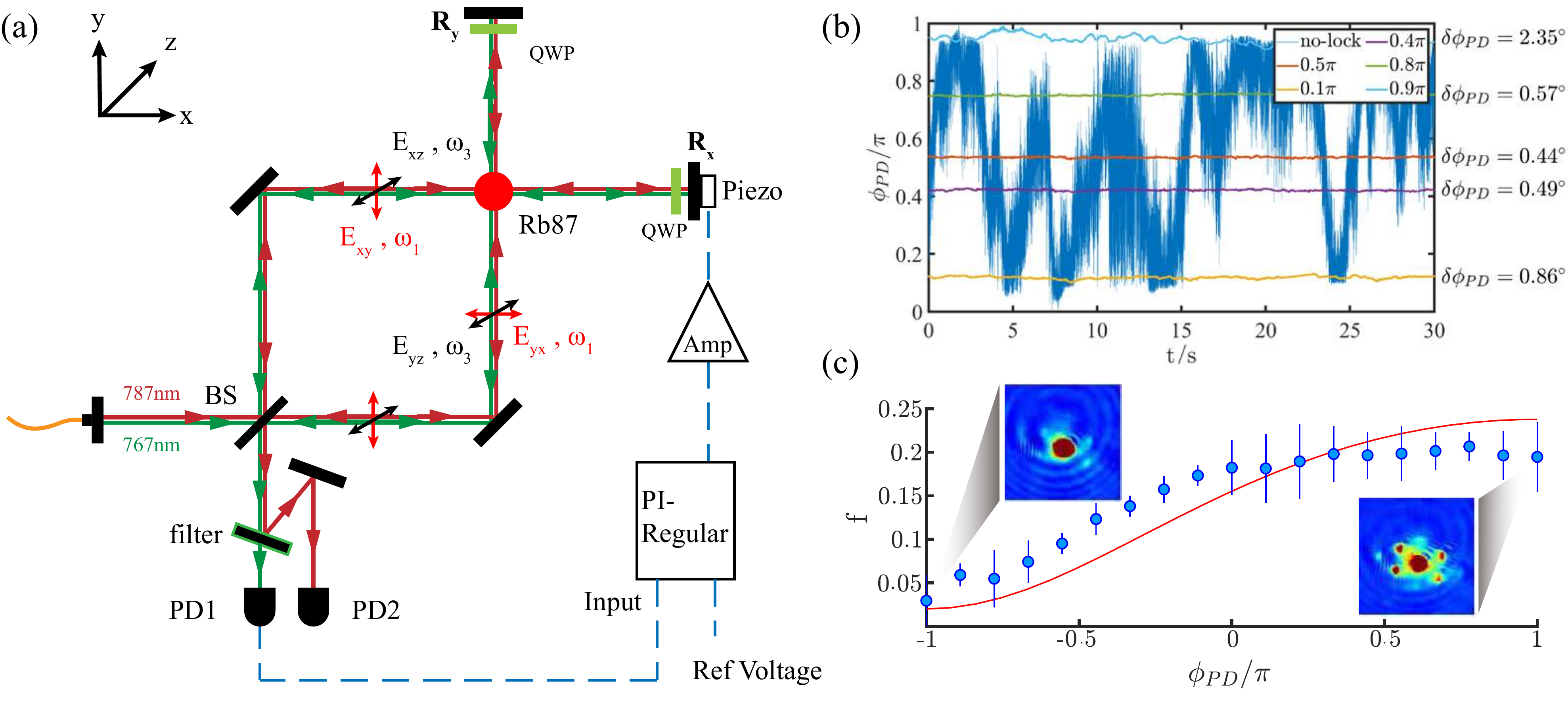}
    \caption{
    (a) Experimental setup for phase lock. (b) The interfered signal of 787nm-laser. The blue lines correspond to the signal without locking, while the other colors (see the figure legends) are signals with phase locking at different reference voltage. The noise of the phase is evaluated by calculating the standard deviation ($\delta\phi_{PD}$) of the signal. (c) The fraction of lattice diffraction VS $\phi_{PD}$. Blue circles with error bars are experimental data. The red line is the theoretical calculation of the fraction.
    }
    \label{PhaseLock}
\end{figure}

\begin{figure}[h]
    \centering
    \includegraphics[width=0.7\linewidth]{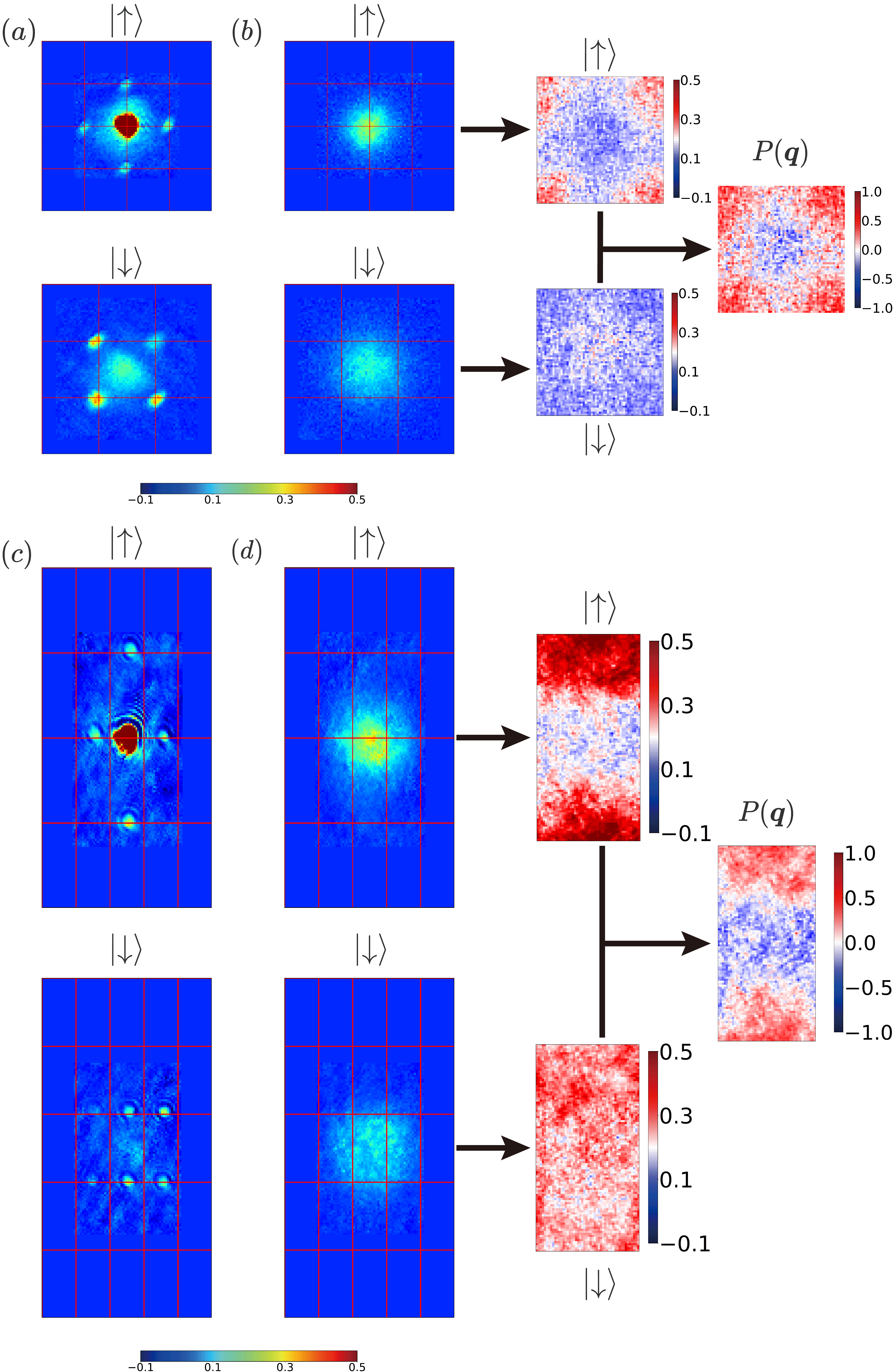}
    \caption{
        Reconstruction of first Brillouin zone for spin texture. (a)-(b): Data obtained on CCD-Z. (a) Ground state for spin up and down. (b): Reconstruction of Spin texture in the FBZ. The size of the FBZ is denoted by the red squares. (c)-(d): Data obtained on CCD-Y.
    }
    \label{FBZ}
\end{figure}

\begin{figure}[h]
    \centering
    \includegraphics[width=0.6\linewidth]{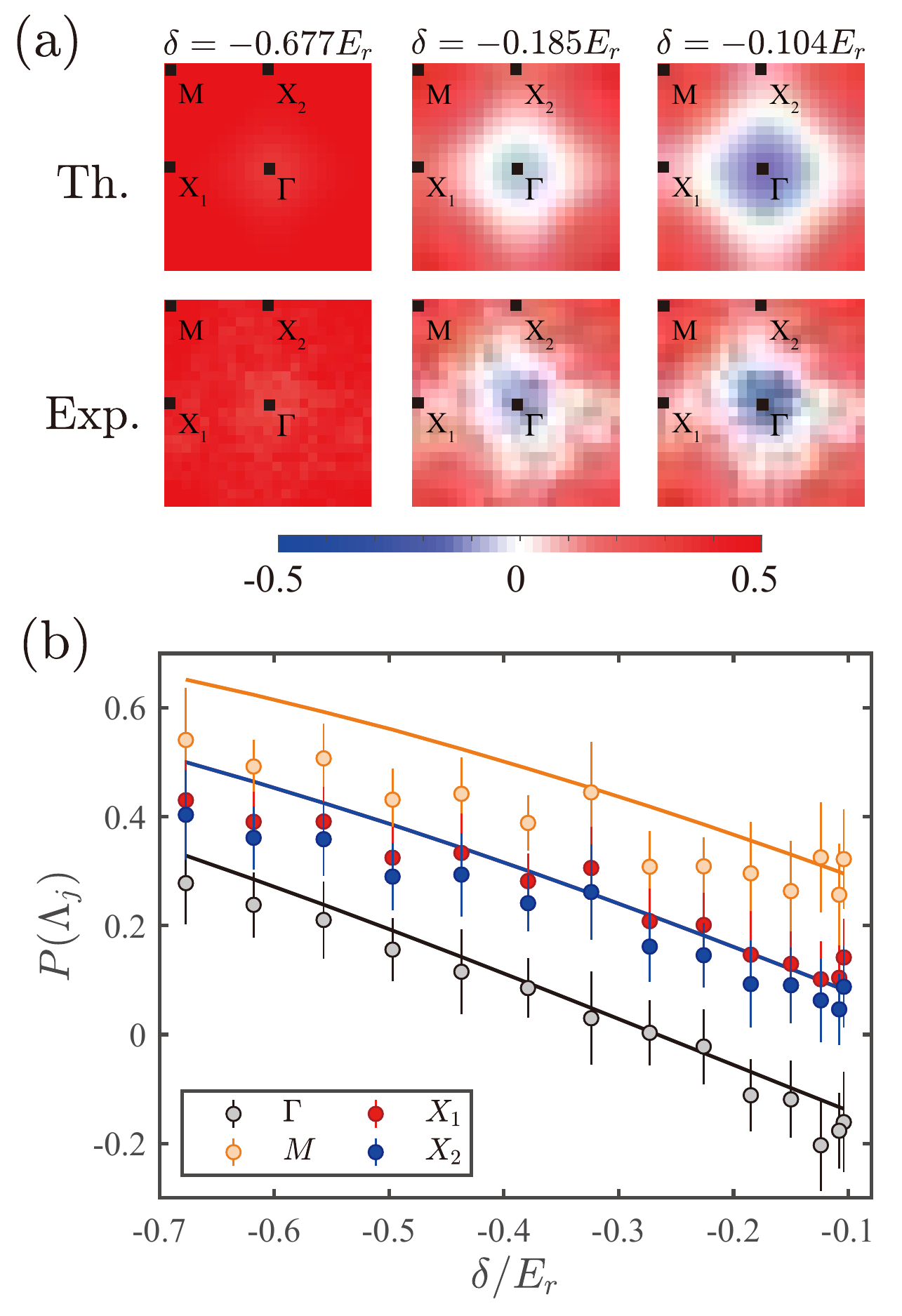}

    \caption{Comparison of experimental with theoretical results for spin textures. (a) 2D Spin textures for the numerical result (up) and the experimental data (down). (b) The spin polarization at high-symmetric momenta with $q_z$ integrated VS $\delta$. The circles with error bars are the experimental data. The lines are numerical results. Both of them are obtained with $V_{2D}=1.77E_r$, $V_{z}=3.54E_r$ and $T=150$nK.
    }
    \label{Virtual}
\end{figure}

\begin{figure}[h]
    \centering
    \includegraphics[width=1\linewidth]{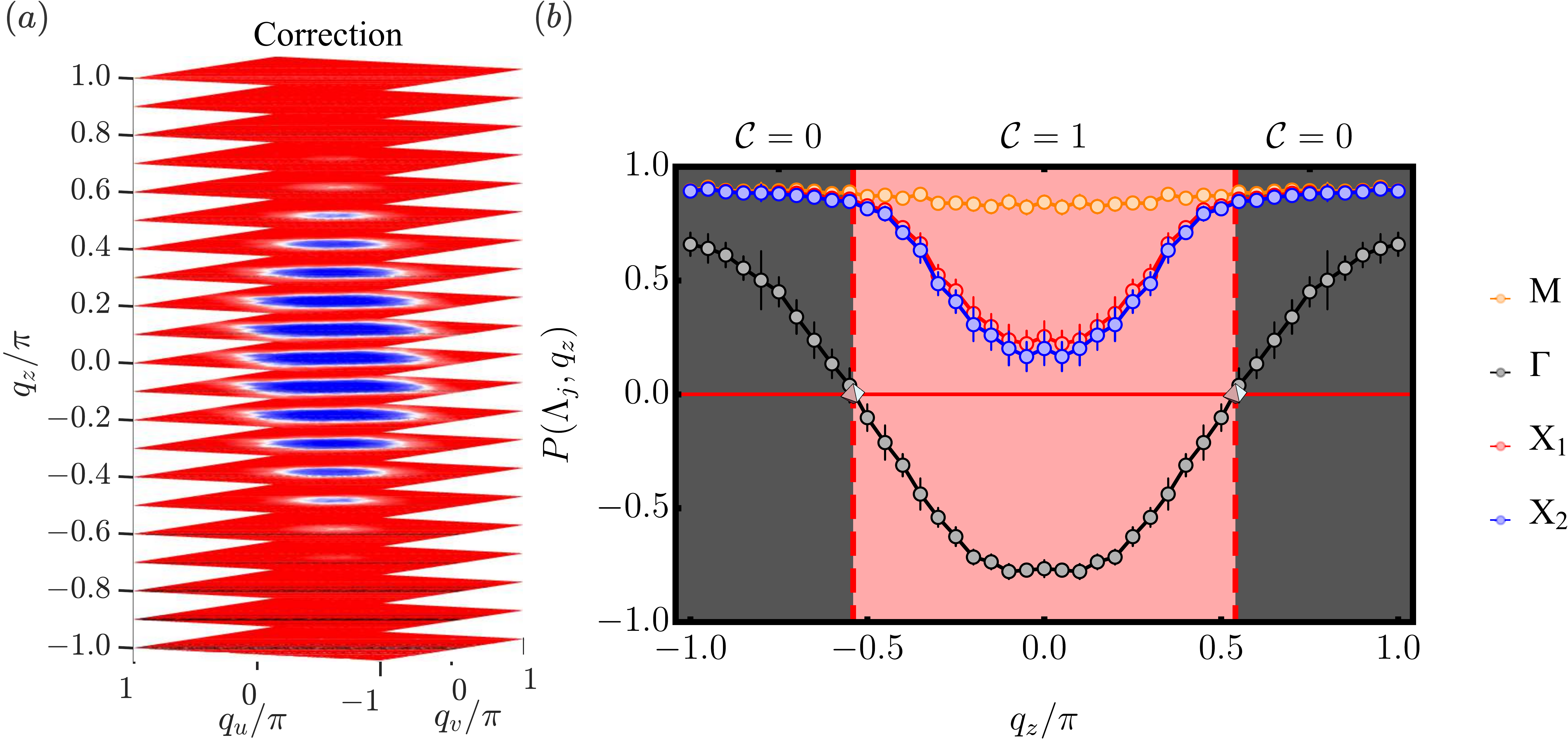}
    \caption{
        High band correction results of \(\delta_0=-0.5E_r\). (a): 2D spin texture stack. (b): Spin polarizations of high symmetric points. Weyl points are marked by diamonds and locate at \(q_z=\pm(0.54\pm0.02)\pi\).
        }
    \label{RemoveHB}
\end{figure}

\begin{figure}
    \centering
    \includegraphics[width=1\linewidth]{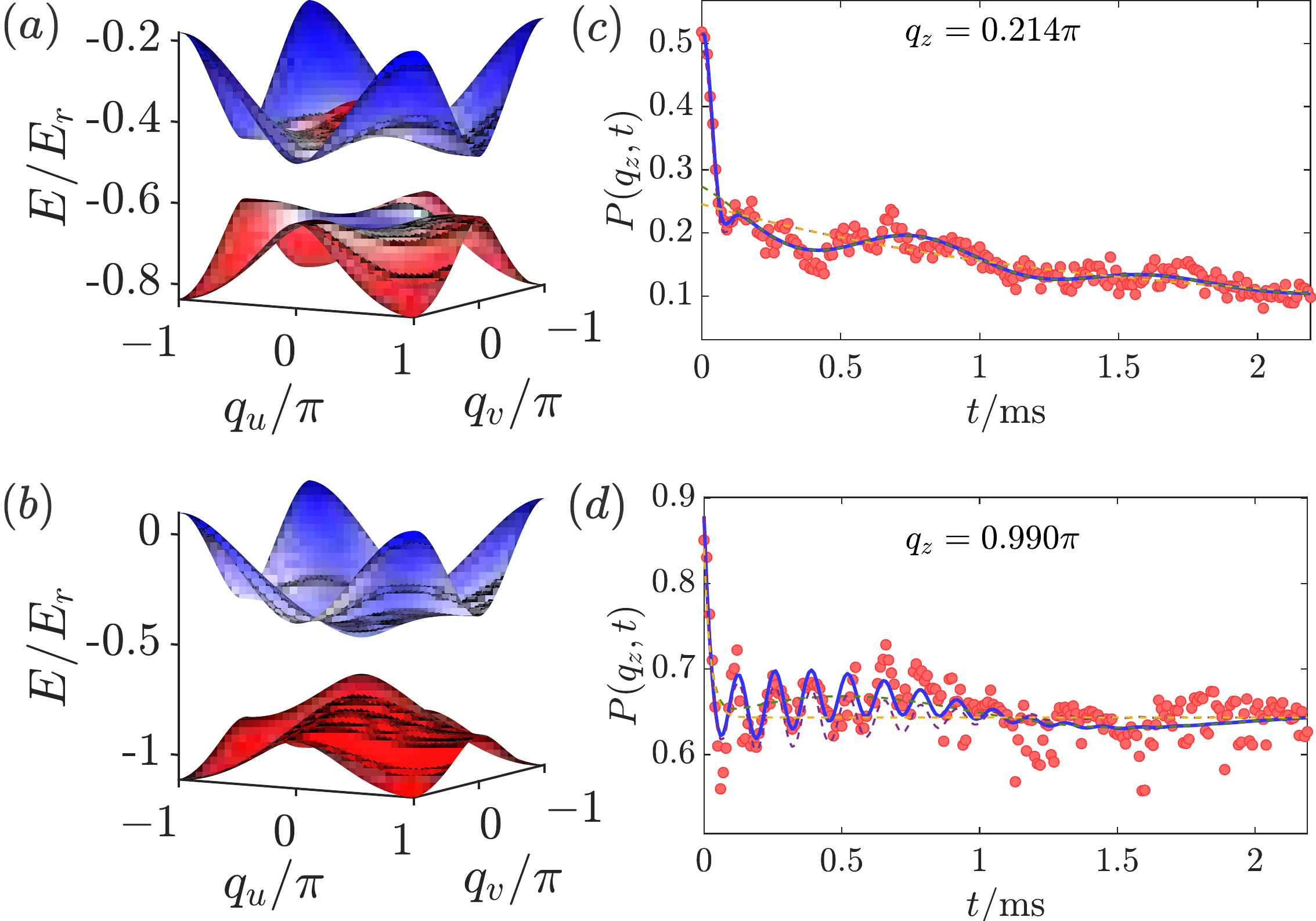}
    \caption{
        Quench dynamics of two typical oscillations of spin polarization. (a)-(b): Band structure on \(q_z=0.214\pi\) plane (a) and \(q_z=0.990\pi\) plane (b). The s bands are topological for (a) and trivial for (b). (c)-(d): Quench dynamics of spin polarization for \(q_z=0.214\pi\) plane (c) and \(q_z=0.990\pi\) plane (d). The experimental data are red circles. Blue solid line: double frequency fitting results. Green dashed line: low frequency component with the background exponential decay. Purple dashed line: high frequency component with the background exponential decay. Orange dashed line: background exponential decay.
    }
    \label{QuenchExp}
\end{figure}

\begin{figure}
    \centering
    \includegraphics[width=1\linewidth]{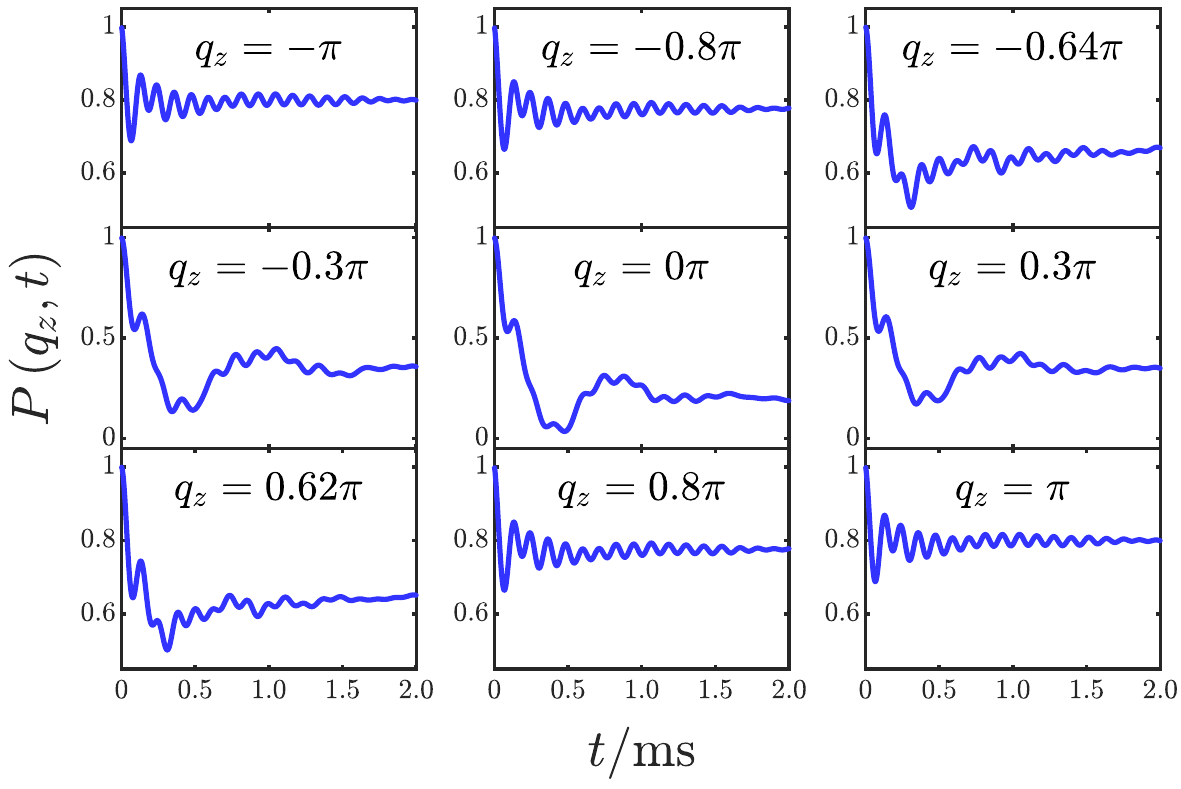}
    \caption{
        Numerical results of evolution of $P(q_z,t)$ for different $q_z$.
    }
    \label{Numeric}
\end{figure}

\setcounter{figure}{0}
\renewcommand{\figurename}{Extended Data Table}
\renewcommand{\thefigure}{\arabic{figure}}

\begin{figure}
    \centering
    \includegraphics[width=1\linewidth]{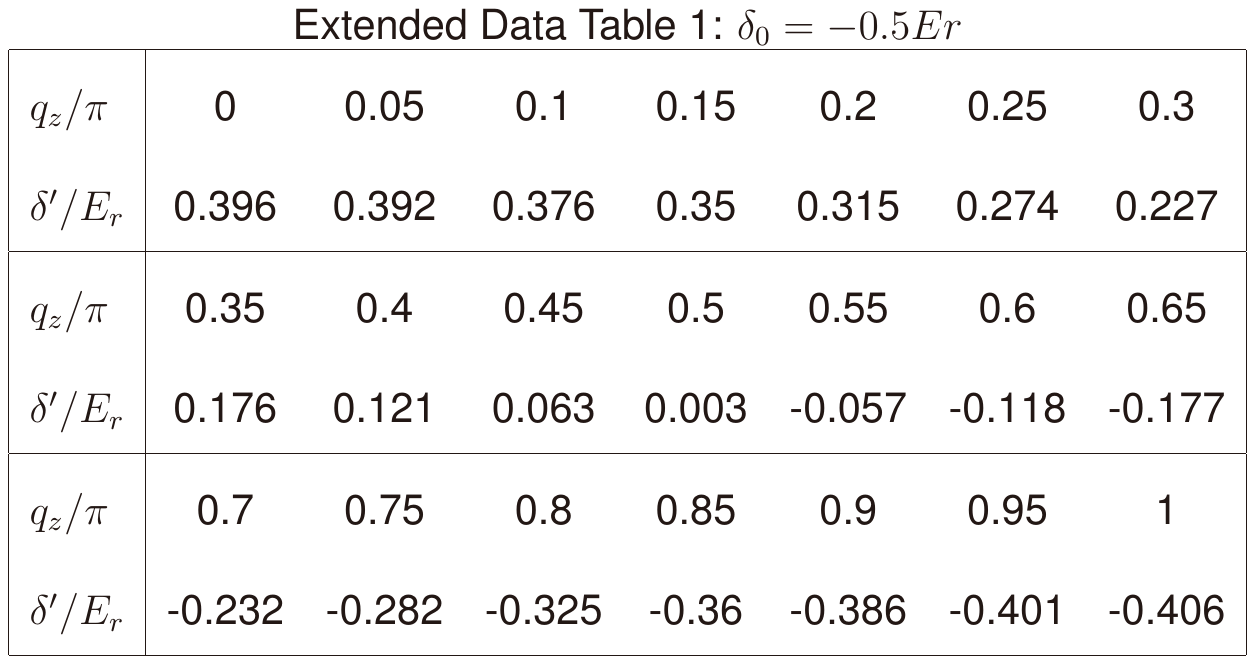}

    \caption{
       Numerical calculations of correspondence between \(\delta^{\prime}\) and \(q_z\) for \(\delta_0=-0.5E_r\).
    }
    \label{Table1}
\end{figure}

\begin{figure}
    \centering
    \includegraphics[width=1\linewidth]{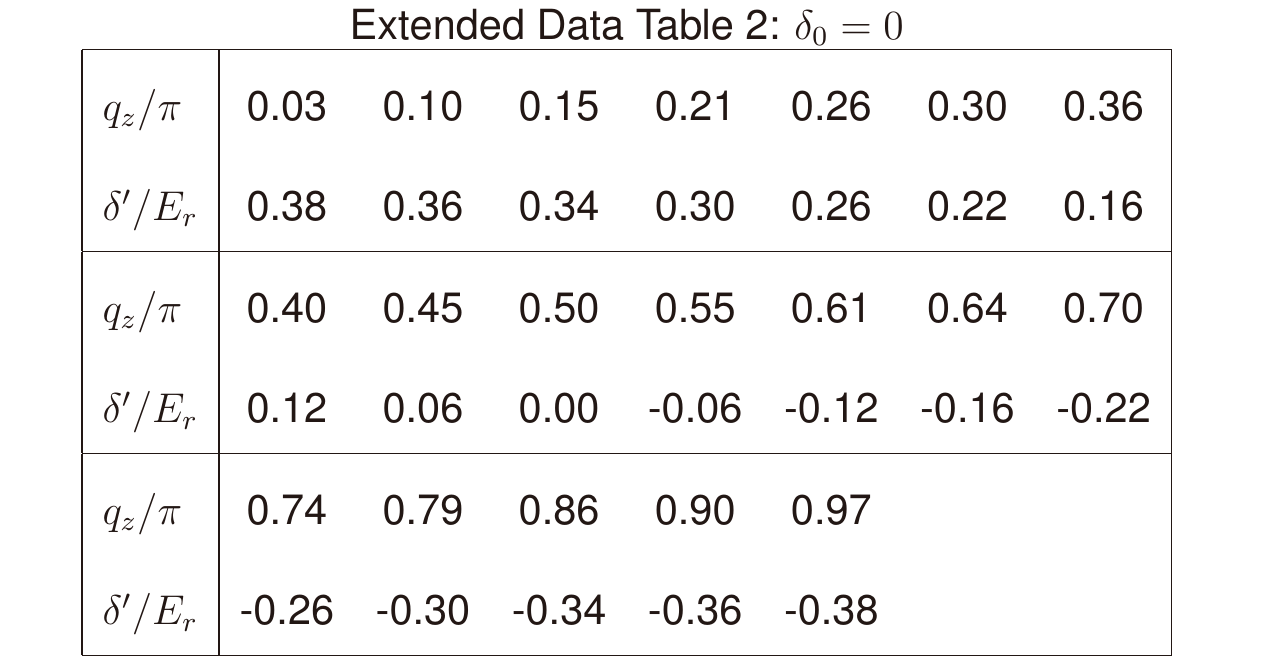}

    \caption{
       Numerical calculations of correspondence between \(\delta^{\prime}\) and \(q_z\) for \(\delta_0=0\).
    }
    \label{Table2}
\end{figure}

\begin{figure}
    \centering
    \includegraphics[width=0.7\linewidth]{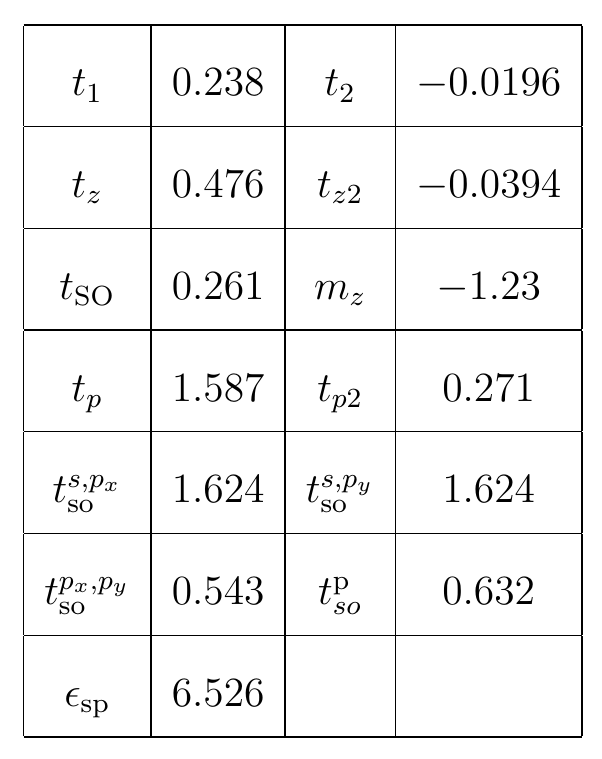}

    \caption{
        Parameters for the TB model.
    }
    \label{Table2}
\end{figure}

\end{document}